\def\CHfour{{\fam0 CH_4}}
\def\CHthree{{\fam0 CH_3}}

\def\HCN{{\fam0 HCN}}
\def\CHtwoOH{{\fam0 CH_2OH}}
\def\CHthreeOH{{\fam0 CH_3OH}}

\def\COtwo{{\fam0 CO_2}}
\def\CO{{\fam0 CO}}
\def\CtwoHtwo{{\fam0 C_2H_2}}
\def\CtwoHthree{{\fam0 C_2H_3}}
\def\CtwoHfour{{\fam0 C_2H_4}}
\def\CtwoHfive{{\fam0 C_2H_5}}
\def\CtwoHsix{{\fam0 C_2H_6}}
\def\CHthreeNHtwo{{\fam0 CH_3NH_2}}
\def\CHtwoNHtwo{{\fam0 CH_2NH_2}}
\def\CHtwoNH{{\fam0 CH_2NH}}

\def\HCO{{\fam0 HCO}}
\def\HtwoCO{{\fam0 H_2CO}}
\def\HtwoCN{{\fam0 H_2CN}}
\def\HtwoO{{\fam0 H_2O}}

\def\Htwo{{\fam0 H_2}}
\def\H{{\fam0 H}}
\def\M{{\fam0 M}}

\def\NH{{\fam0 NH}}
\def\NHtwo{{\fam0 NH_2}}
\def\NHthree{{\fam0 NH_3}}
\def\Net{{\fam0 Net:}}
\def\N{{\fam0 N}}
\def\Ntwo{{\fam0 N_2}}
\def\NtwoH{{\fam0 N_2H}}

\def\NtwoHtwo{{\fam0 N_2H_2}}

\def\OH{{\fam0 OH}}

\def\O{{\fam0 O}}

\def\cmtwo{{\fam0 cm^2}}

\def\scinot#1.{\hbox{$\times 10^{#1}$}}
\def\smone{{\fam0\,s^{-1}}}

\def\ten#1.{\hbox{$10^{#1}$}}

\def\deg{\ifmmode^\circ\else$\null^\circ$\fi}
\def\spose#1{\hbox to 0pt{#1\hss}}
\def\lta{\mathrel{\spose{\lower 3pt\hbox{$\mathchar "218$}}\raise 2.0pt\hbox{$\mathchar"13C$}}}
\def\gta{\mathrel{\spose{\lower 3pt\hbox{$\mathchar "218$}}\raise 2.0pt\hbox{$\mathchar"13E$}}}
\def\lrarrow{\mathrel{\spose{\lower 1pt\hbox{$\rightarrow$}}\raise 3.0pt\hbox{$\leftarrow$}}}

\newcommand{\teff}{$T_{\footnotesize\textsl{eff}}$}
\newcommand{\kdeep}{$K_{\footnotesize\textsl{deep}}$}
\newcommand{\kzz}{$K_{\footnotesize\textsl{zz}}$}

\documentclass{emulateapj}

\slugcomment{submitted to Astrophys. J.}

\shorttitle{The Composition of Directly Imaged Giant Planets}
\shortauthors{Moses et al.}

\begin{document}


\title{On the Composition of Young, Directly Imaged Giant Planets}

\author{J. I. Moses}
\affil{Space Science Institute, 4750 Walnut Street, Suite 205, Boulder, CO 80301, USA}
\email{jmoses@spacescience.org}

\author{M. S. Marley} \author{K. Zahnle}
\affil{NASA Ames Research Center, Moffett Field, CA 94035, USA}

\author{M. R. Line} \author{J. J. Fortney}
\affil{Department of Astronomy and Astrophysics, University of California, Santa Cruz, CA 95064, USA}

\author{T. S. Barman}
\affil{Lunar and Planetary Laboratory, University of Arizona, Tucson, AZ 85721, USA}

\author{C. Visscher}
\affil{Dordt College, Sioux Center, IA 51250, USA}
\affil{Space Science Institute, Boulder, CO 80301, USA}

\author{N. K. Lewis}
\affil{Space Telescope Science Institute, Baltimore, MD 21218, USA}

\and

\author{M. J. Wolff}
\affil{Space Science Institute, Boulder, CO 80301, USA}

%
%

\begin{abstract}
The past decade has seen significant progress on the direct detection and characterization of young, self-luminous 
giant planets at wide orbital separations from their host stars. Some of these planets show evidence for disequilibrium 
processes like transport-induced quenching in their atmospheres; photochemistry may also be important, despite the
large orbital distances. These disequilibrium chemical processes can alter the expected composition, spectral behavior, 
thermal structure, and cooling history of the planets, and can potentially confuse determinations of bulk elemental 
ratios, which provide important insights into planet-formation mechanisms. Using a thermo/photochemical kinetics and 
transport model, we investigate the extent to which disequilibrium chemistry affects the composition and spectra of 
directly imaged giant exoplanets. Results for specific ``young Jupiters'' such as HR 8799 b and 51 Eri b are 
presented, as are general trends as a function of planetary effective temperature, surface gravity, incident ultraviolet 
flux, and strength of deep atmospheric convection. We find that quenching is very important on young Jupiters, leading 
to CO/CH$_4$ and N$_2$/NH$_3$ ratios much greater than, and H$_2$O mixing ratios a factor of a few less than, 
chemical-equilibrium predictions. Photochemistry can also be important on such planets, with CO$_2$ and HCN being key 
photochemical products. Carbon dioxide becomes a major constituent when stratospheric temperatures are low and recycling 
of water via the $\Htwo$ + OH reaction becomes kinetically stifled.  Young Jupiters with effective temperatures $\lta$700 K 
are in a particularly interesting photochemical regime that differs from both transiting hot Jupiters and our own 
solar-system giant planets. 
\end{abstract}



\section{Introduction}

Most of the exoplanets discovered to date have been identified through transit observations or 
radial-velocity measurements --- techniques that favor the detection of large planets orbiting close 
to their host stars.  Direct detection of a planet within the overwhelmingly glare and 
non-negligible point-spread function of its brighter star is challenging and requires high-contrast observations, 
often with adaptive-optics techniques from large telescopes on the ground or in space.  As a result of these 
observational challenges, direct imaging favors the detection of massive, self-luminous (i.e., young) giant planets 
at wide orbital separations from their host stars.  These ``young Jupiters'' are hot at depth because the 
leftover accretional and gravitational potential energy from the planet's formation has not had time to 
convect up through the atmosphere and be radiated away yet.  Only $\sim$3\% of the currently confirmed 
exoplanets\footnote{See http://exoplanet.eu, http://{\-}exoplanetarchive.{\-}ipac.{\-}caltech. edu, 
or http://{\-}www.{\-}openexoplanetcatalogue.{\-}com} have been detected through direct imaging, but these 
planetary systems have high intrinsic interest.  For example, they serve as potential analogs to our own solar system 
in its formative years,
when Jupiter and our other giant planets were born 
and evolved behind ice condensation fronts in the solar nebula but never migrated inward --- unlike, apparently, 
many of the known close-in, transiting, extrasolar giant planets.  Directly imaged planets therefore provide a window 
into our own past and provide important clues to our solar system's origin and evolution \citep[see, e.g.,][]{madhu14rev}.
Wavelength-dependent photometry or spectra of directly imaged planets can also provide useful constraints on 
atmospheric properties such as composition, thermal structure, metallicity, bulk elemental ratios, and 
the presence or absence of clouds \citep[see the reviews of][]{madhu14rev,bailey14rev,crossfield15,madhu16rev}.


Short-period, transiting ``hot Jupiters'' and directly imaged ``young Jupiters'' both have similar effective 
temperatures, often ranging from $\sim$500 to 2500 K. However, in terms of their thermal structure and spectral 
appearance, directly imaged planets have more in common with brown dwarfs than with hot Jupiters 
\citep[e.g.,][]{burrows03,fort08colors,bowler16}.  In particular, the stratospheres (radiative regions) of directly 
imaged planets and brown dwarfs are much cooler than those of highly-irradiated hot Jupiters, and the cooler 
regions overlying hot continuum regions at depth can result in potentially deeper molecular absorption bands being 
present in emission spectra \citep{madhu14rev}.  It can therefore be easier to detect atmospheric molecules on young 
Jupiters and brown dwarfs, unless high clouds are present to obscure the absorption features.

One drawback of direct imaging is that the planet's radius and mass cannot be well determined, unlike the situation 
with, respectively, transit observations and radial-velocity measurements.  Instead, the mass and radius of directly 
imaged planets are more loosely constrained through atmospheric modeling and comparisons with the observed luminosity 
and spectral/photometric behavior, often in combination with estimates of the age of the system and constraints from 
evolutionary models.  The theoretical modeling and model-data comparisons can result in degeneracies between the planet's 
apparent size, surface gravity, effective temperature, and cloud properties
\citep[e.g.,][]{marley07,marley12,barman11hr8799,barman112mass,barman15,currie11,madhu11hr8799,spiegel12,bonnefoy13,bonnefoy16,lee13,skemer14,baudino15,morzinski15,zurlo16}.  

On the other hand, the identification of molecular features in the observed spectra is typically unambiguous
on young Jupiters \citep[e.g.,][]{konopacky13,barman15}, and H$_2$O, CO, and/or CH$_4$ have been detected in
in spectra from several directly imaged planets
\citep{patience10,barman11hr8799,barman112mass,barman15,oppenheimer13,konopacky13,janson13,snellen14,chilcote15,macintosh15}. 
The apparent deficiency of methane features on many cooler directly imaged planets, in conflict with chemical equilibrium 
expectations, has been argued as evidence for disequilibrium processes like transport-induced quenching on these 
planets
\citep[e.g.,][]{bowler10,hinz10,janson10,janson13,barman11hr8799,barman112mass,barman15,galicher11,marley12,skemer12,skemer14,ingraham14,currie14,zahnle14},
and so many of the above groups included quenching in their theoretical modeling \citep[see][for more details about CO $\leftrightarrow$ 
CH$_4$ quenching on directly imaged planets and brown dwarfs]{visscher11,zahnle14}.
Other disequilibrium chemical processes such as photochemistry have been assumed to be unimportant due to the large orbital 
distances of these planets \citep{crossfield15}; however, the young stellar hosts of directly imaged planets tend to be bright 
in the ultraviolet, making photochemistry potentially important \citep[e.g.,][]{zahnle16}.

The goal of the present investigation is to quantify the extent to which disequilibrium chemical processes like photochemistry 
and quenching affect the composition and spectra of young, directly imaged planets.  Our main theoretical tool is a 
thermochemical-photochemical kinetics and transport model \citep[e.g.,][]{moses11,visscher11,moses13coratio,moses13gj436} 
that tracks the chemical production, loss, and transport of the most abundant gas-phase species in a hydrogen-dominated 
planetary atmosphere.  We calculate the expected composition of specific directly imaged exoplanets such as 51 Eri b 
and HR 8799 b,
for which observational spectra are available, as well as investigate how the composition of generic 
``young Jupiters'' is affected by planetary parameters such as 
the effective temperature, surface gravity, incident ultraviolet flux, and the strength of atmospheric mixing.
We also explore how disequilibrium chemistry affects the resulting spectra of directly imaged planets.

\section{Theoretical Model}


To calculate the vertical profiles of atmospheric species on directly imaged planets, we use the 
Caltech/JPL KINETICS code \citep{allen81,yung84} to solve the coupled one-dimensional (1D) continuity 
equations for 92 neutral carbon-, oxygen-, nitrogen-, and hydrogen-bearing species that interact through 
$\sim$1650 kinetic reactions.  Hydrocarbons with up to six carbon atoms are considered, although the 
reaction list becomes increasingly incomplete the heavier the molecule.  We do not consider ion chemistry 
from photoionization \citep{lavvas14alkali} or galactic-cosmic-ray ionization \citep{rimmer14}.  Ion 
chemistry is not expected to affect the mixing ratios of the dominant gas species, but it will likely augment 
the production of heavy organic molecules, just as on Titan \citep[e.g.,][]{waite07,vuitton07}.  
Lacking any definitive evidence to the contrary for directly imaged giant planets, we assume the 
atmospheres have a solar elemental composition.

The reaction list includes both ``forward'' (typically exothermic) reactions and their reverses, where the reverse 
reaction rate coefficient is calculated from the forward rate coefficient and equilibrium constant assuming 
thermodynamic reversibility \citep[e.g.,][]{visscher11,heng16}.  All reactions except those involving photolysis 
are reversed.  The fully reversed reaction mechanism ensures that thermochemical equilibrium is maintained kinetically 
in the hotter deep atmosphere, while disequilibrium photochemistry and transport processes can take over and dominate 
in the cooler upper atmosphere 
\citep[e.g.,][]{moses11,line11gj436,venot12,koppa12,miller-ricci12,agundez14pseudo,miguel14,hu14,benneke15,zahnle16}.  
The model automatically accounts for the transport-induced quenching of species, whereby mixing ratios are 
``frozen in'' at a constant mixing ratio above some quench pressure as vertical transport processes start to 
dominate over the chemical reactions that are attempting to drive the atmosphere back toward thermochemical 
equilibrium \citep{prinn77,lewis84,fegley94}.  

The quenching process depends on the adopted reaction mechanism 
\citep[cf.][]{visscher10co,moses11,visscher11,line11gj436,venot12,moses14,zahnle14,wang15deep,rimmer16}. 
Our chemical reaction list is taken from \citet{moses13gj436} and includes a thorough review of the 
key reaction mechanisms of potential importance in the quenching of CO $\leftrightarrows$ CH$_4$ and 
N$_2$ $\leftrightarrows$ NH$_3$ \citep{visscher10co,visscher11,moses10,moses11,moses13coratio,moses13gj436,moses14};  
further details of the thermo/photochemical kinetics and transport model are provided in the above papers, and the 
reaction list is provided in the journal supplementary material.  Note that we do not include the fast rate coefficient 
for H + CH$_3$OH $\rightarrow$ CH$_3$ + H$_2$O suggested by \citet{hidaka89} that is controlling CO-CH$_4$ quenching 
in the \citet{venot12} mechanism.  As discussed by \citet{norton90}, \citet{lendvay97}, and \citet{moses11}, this 
reaction actually possesses a very high energy barrier ($>$ 10,000 K) and is not expected to be important under either 
methanol-combustion conditions or in the deep atmospheres of hydrogen-rich exoplanets --- in other words, the Hidaka 
et al.~rate coefficient greatly overestimates the rate of this reaction.  \citet{zahnle14} adopt the upper limit for 
this reaction as suggested by \citet{norton89} and find it to be important but not typically dominant in CO--CH$_4$ 
quenching, except for cooler brown dwarfs with weak mixing.  We adopt the much smaller rate coefficient as calculated 
by \citet{moses11}, and this reaction does not play a role in CO--CH$_4$ quenching.  Similarly, we do not adopt the 
relatively fast rate-coefficient expression for NH$_2$ + NH$_3$ $\rightarrow$ N$_2$H$_3$ + H$_2$ estimated by \citet{konnov00} 
that is affecting N$_2$-NH$_3$ quenching in the \citet{venot12} mechanism, as again, this reaction is expected to have a 
high-energy barrier and be slower under relevant conditions than the Konnov and De Ruyck estimate \citep[e.g.,][]{dean84}.

Our model grids consist of 198 vertical levels separated uniformly in log(pressure) (providing multiple grid
levels per scale height to insure accurate diffusion calculations), with a bottom level 
defined where the deep atmospheric temperature on an adiabatic gradient is greater than $\sim$2700 K 
(to insure that the N$_2$-NH$_3$ quench point is captured), and a 
top level residing at $\sim$10$^{-8}$ mbar (to insure all the molecular absorbers are optically thin in the 
ultraviolet).  The top region of our model grid extends through what would 
typically be the ``thermosphere'' of the planet; however, we neglect non-stellar sources of 
thermospheric heating (such as auroral and Joule heating), which are poorly understood but are 
important on our solar-system 
giant planets \citep[e.g.,][]{yelle04book,nagy09}.  Our results should therefore only be considered reliable from 
the deep troposphere on up to the homopause level at the base of the thermosphere (near 10$^{-4}$ to 10$^{-6}$ mbar, 
depending on the strength of atmospheric mixing), where molecular diffusion acts to limit the abundance of heavy 
molecular and atomic species in the lighter background hydrogen atmosphere.  

\begin{figure}
\includegraphics[scale=0.44]{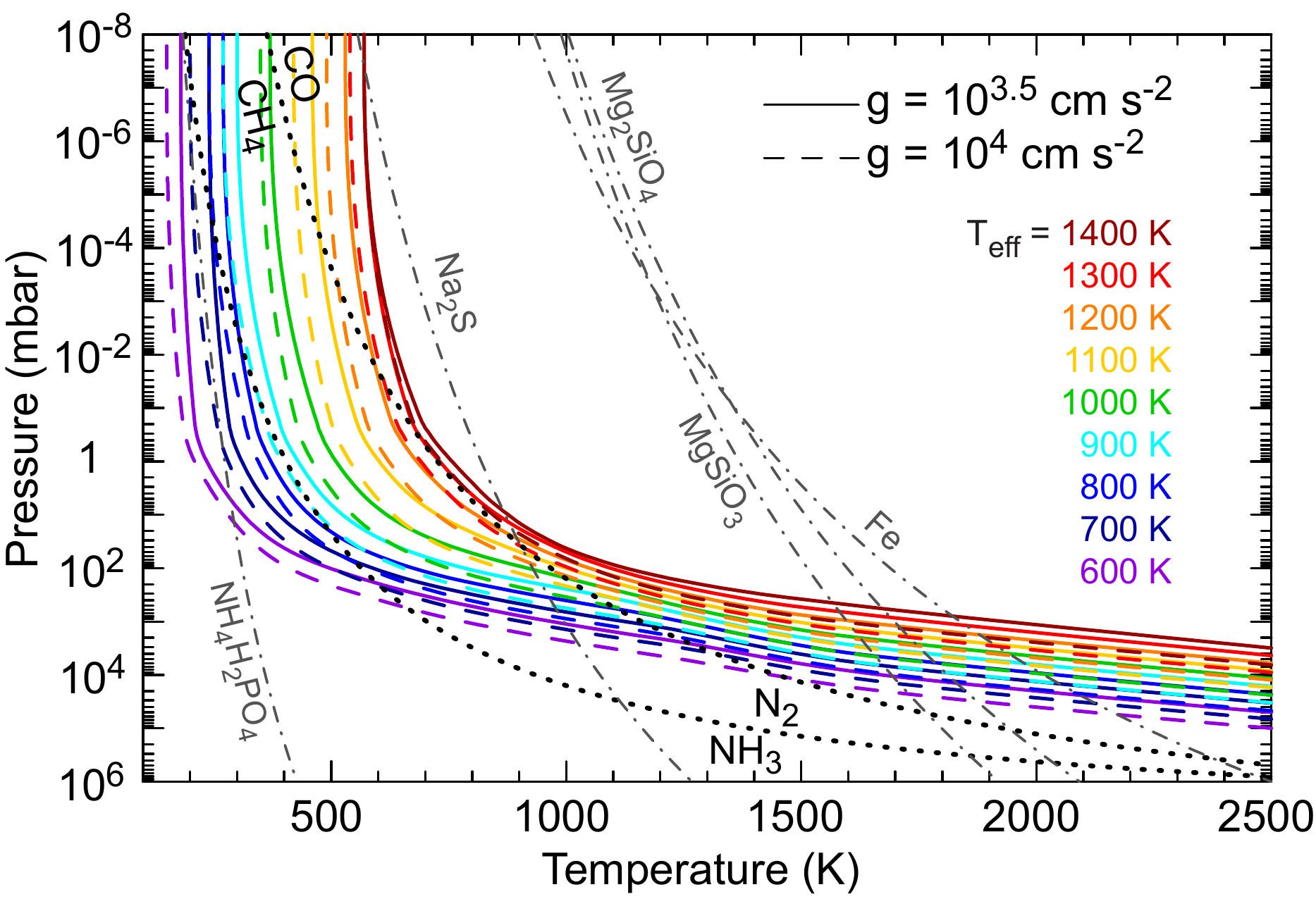}
\caption{Theoretical temperature profiles for generic directly imaged planets from the radiative-convective 
equilibrium model of \citet{marley12}, as a function of effective temperature $T_{\footnotesize\textsl{eff}}$ for an assumed 
surface gravity (in cm s$^{-2}$) of $\log(g)$ = 3.5 (colored solid lines) and $\log(g)$ = 4.0 (colored dashed 
lines) and assumed solar composition atmosphere in chemical equilibrium.  Profiles are shown every 100 K from 
$T_{\footnotesize\textsl{eff}}$ = 600 to 1400 K.  The gray dot-dashed lines show the condensation curves for some important 
atmospheric cloud-forming species (as labeled) for an assumed solar-composition atmosphere.  The thicker dotted 
black lines represent the boundaries where CH$_4$ and CO have equal abundances and where N$_2$ and NH$_3$ have 
equal abundance in chemical equilibrium for solar-composition models.  Methane and ammonia dominate to the lower 
left of these curves, while CO and N$_2$ dominate to the upper right.  Note that all the profiles remain within 
the CO-dominated regime at depth, whereas all except for the hottest planets transition to the CH$_4$-dominated 
regime at higher altitudes.
A color version of this figure is available in the online journal.\label{figtemp}}
\end{figure}

The thermal structure itself is not 
calculated self-consistently but is adopted from two different atmospheric models: (1) the radiative-convective 
equilibrium models described in \citet{mckay89}, \citet{marley99}, \citet{marley02}, and \citet{saumon08}, with 
updates as described in \citet{marley12}, and (2) the PHOENIX-based models described in \citet{hauschildt97}, 
\citet{allard01}, and \citet{barman11hr8799}, with updates as described in \citet{barman15}.  We add a smoothly 
varying, nearly isothermal profile at the top of the above-mentioned theoretical model profiles to extend our 
grids to lower pressures, except in isolated cases where we test the effects of a hotter (1000 K) thermosphere. 
Figure~\ref{figtemp} shows the temperature profiles adopted for our cloud-free generic 
directly imaged planets, as a function of effective temperature $T_{\small\textsl{eff}}$ 
for two different assumed 1-bar surface gravities, $\log(g)$ = 3.5 and 4.0 cgs.  These profiles are 
calculated without considering stellar irradiation --- for all directly imaged planets discovered to date, 
the external radiation field has little effect on the thermal profile due to the planets' large orbital distance 
and strong internal heat flux.
As such, the internal heat completely dominates the thermal structure, and temperatures on these planets are 
hotter at depth and colder in the stratospheric radiative region than for close-in transiting giant planets of the 
same effective temperature. The profiles from Fig.~\ref{figtemp} were generated with the NASA Ames brown-dwarf and 
exoplanet structure models \cite[e.g.,][]{marley12}; tables with the individual pressure-temperature structure from 
these models can be found in the journal Supplementary Material.
Disequilibrium processes like photochemistry and quenching are expected to have a relatively minor effect on 
the thermal structure \citep[e.g.,][]{agundez14gj436}, unless these processes affect the H$_2$O abundance.  

Given a temperature-pressure profile, the NASA CEA code of \citet{gordon94} is then used to determine the 
chemical-equilibrium abundances, which are used as initial conditions in the photochemical model.  We use the protosolar 
abundances listed in Table 10 of \citet{lodders10} to define our ``solar'' composition.  
The mean molecular mass profile from the chemical-equilibrium solution, the pressure-temperature profile, and the assumed 
physical parameters of the planet become inputs to the hydrostatic equilibrium equation, whose solution sets the altitude 
scale and other atmospheric parameters along the vertical model grid.  For a surface (1-bar) gravity of $g$ = 10$^4$ cm s$^{-2}$, 
the planet mass M$_p$ is $4 M_J$, and for $g$ = 10$^{3.5}$ cm s$^{-2}$, M$_p$ = $2 M_J$.  For boundary conditions, we assume 
the fluxes of the species are zero at the top and bottom of the model.  The models are run until steady state, with a 
convergence criterion of 1 part in 1000.  For the photochemical calculations, the atmospheric extinction is calculated 
from the absorption and multiple Rayleigh scattering of gases only --- aerosol extinction is ignored due to a lack of 
current predictive capability regarding the hazes.  The atmospheric radiation field for the photochemical model is 
calculated for diurnally averaged conditions for an assumed (arbitrary) 24-hour rotation period at 30\deg\ latitude at 
vernal equinox, with an assumed zero axial tilt for the planet.  These assumptions should provide acceptable ``global 
average'' conditions for most young Jupiters.

As is standard in 1D photochemical models, we assume that vertical transport occurs through molecular and 
``eddy'' diffusion, with the eddy diffusion coefficient profile \kzz$(z)$ being a free parameter.  
The molecular diffusion coefficients assumed in the model are described in \citet{moses00a}.  Although 
vertical transport of constituents in real atmospheres occurs through convection, large-scale advection, 
atmospheric waves, and turbulent ``eddies'' of all scales, this constituent transport often mimics diffusion 
\citep{lindzen81,strobel81,brasseur99}, and the concept of eddy diffusion has proven to be a useful one for 
atmospheric models.  The eddy diffusion coefficient profile for an atmosphere cannot typically be derived 
accurately from first principles.  Instead, observations of chemically long-lived species are used to empirically 
constrain \kzz$(z)$ \citep[e.g.,][]{allen81,atreya84,moses05}.  On H$_2$-dominated planets and brown dwarfs, 
the relative abundance of CO and CH$_4$ can be used to constrain \kzz\ at the quench point 
\citep[see][]{prinn77,fegley94,griffith99,visscher11}.  For most directly imaged planets planets, the CO-CH$_4$ quench 
point will reside in the deep, convective portion of the atmosphere, where free-convection and mixing-length 
theories \citep[e.g.,][]{stone76} predict relatively large eddy diffusion coefficients and rapid mixing 
(e.g., \kzz\ $\gta$ 10$^{10}$ $\cmtwo$ $\smone$ for most young Jupiters, assuming the atmospheric scale 
height as the mixing length).  However, the mixing length to use for these expressions is not obvious \citep{smith98,freytag10}, 
and the quench point for some planets may approach the radiative region, where \kzz\ is expected to drop off 
significantly before increasing roughly with the inverse square root of atmospheric pressure due to the action 
of atmospheric waves \citep[e.g.,][]{lindzen81,strobel81,parmentier13}.  

\begin{figure}
\includegraphics[scale=0.44]{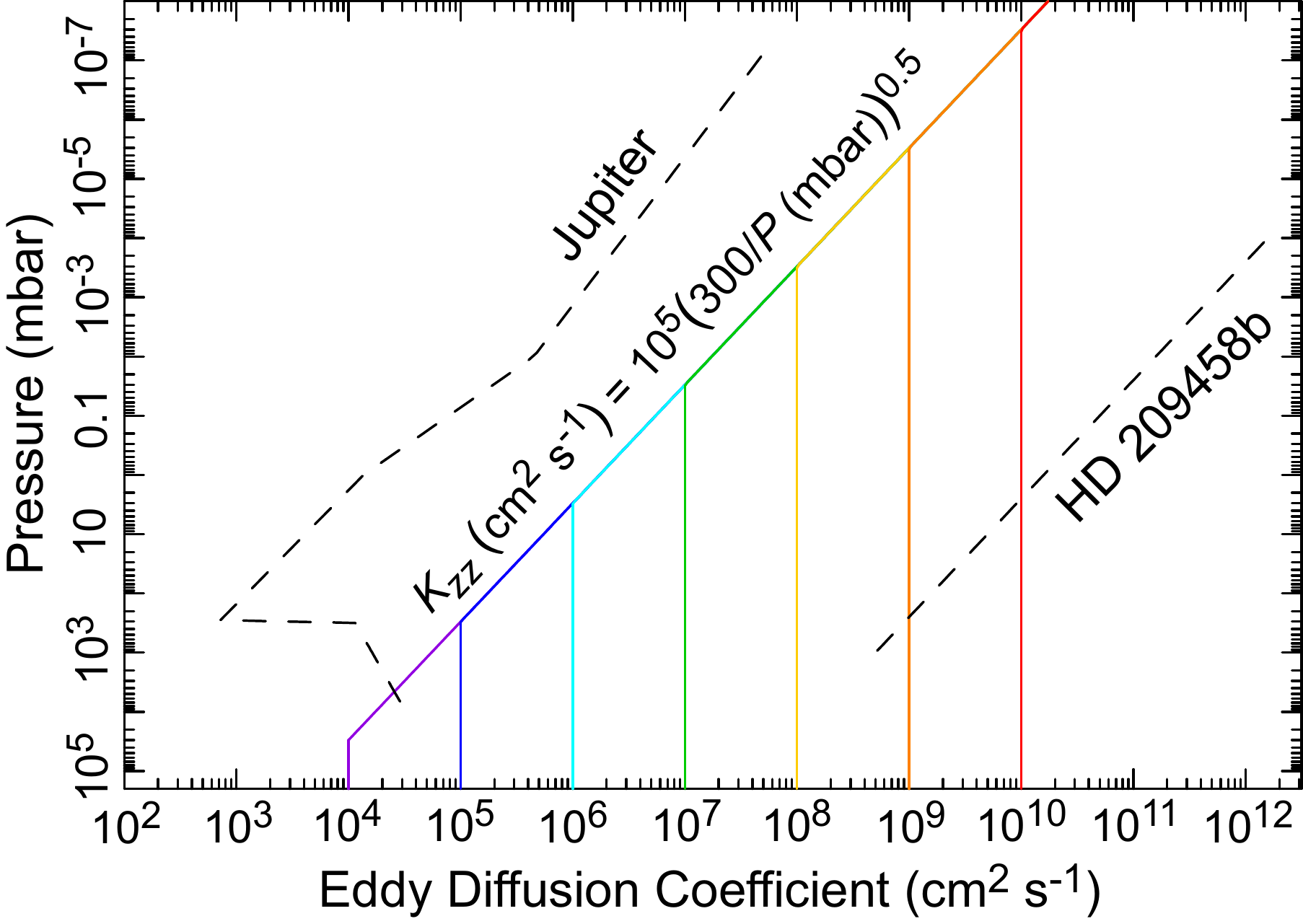}
\caption{Eddy diffusion coefficient profiles (colored solid lines) adopted in our thermo/photochemical kinetics 
and transport models.  The \kzz\ profiles are assumed to vary as 10$^5 \left( 300/P_{\small\textrm{mbar}} 
\right)^{0.5}$ cm$^2$ s$^{-1}$ in the radiative region, with different models having different cutoff values 
($K_{\footnotesize\textsl{deep}}$) at depth.  Profiles derived for Jupiter \citep{moses05} and the hot 
Jupiter HD 209458b \citep{parmentier13} are shown for comparison (dashed lines).
A color version of this figure is available in the online journal.\label{figeddy}}
\end{figure}

We therefore explore a range of possible \kzz\ profiles, with roughly constant values at depth, trending to values that 
vary as 1/$\sqrt{P}$ as the pressure $P$ decreases.  In particular, we assume that \kzz\ (cm$^2$ s$^{-1}$) = 10$^5 
\left( 300/P_{\small\textrm{mbar}} \right)^{0.5}$ in the radiative region above $\sim$300 mbar 
(hereafter called the stratosphere), but we do not let \kzz\ drop below some value ``$K_{\footnotesize\textsl{deep}}$'' 
that varies with the different models considered (see Fig.~\ref{figeddy}).  This convention allows the different 
models for a given \teff\ to have a similar homopause pressure level in the upper atmosphere (i.e., the pressure 
level to which the molecular species can be mixed before molecular diffusion starts to limit their abundance), 
while still testing the effect of variations in \kzz\ at the quench point.   

Note from Fig.~\ref{figeddy} that we have chosen stratospheric \kzz\ profiles that are intermediate between those 
derived empirically from chemical tracers for our own solar-system (cold) Jupiter \citep{moses05} and those 
derived from tracer transport in 3D dynamical models of the hot transiting exoplanet HD 209458b \citep{parmentier13}, 
which seems reasonable given that atmospheric temperatures for directly imaged planets are intermediate between the 
two.  Eddy diffusion coefficients scale directly with vertical velocities and atmospheric length scales, and
both tend to be larger for higher temperatures.  Young Jupiters are very hot and convective at depth, but their 
stratospheres are relatively cold and statically stable.

When estimating \kzz\ profiles for exoplanetary atmospheres, we keep in mind that atmospheric waves are 
typically responsible for mixing in the stratosphere \citep[e.g.,][]{lindzen81}, and wave activity could be 
correlated with both the strength of stellar insolation and internal heat, as the main drivers for these waves.  
In the troposphere, convection dominates, and mixing is stronger for higher internal heat fluxes.  For example, 
in the \citet{freytag10} hydrodynamic models of cool dwarfs, the maximum effective tropospheric diffusion coefficient 
(analogous to our ``\kdeep'') increases with increasing \teff\ over the whole 900 $\le$ \teff\ $\le$ 2800 K 
model range examined.  \citet{freytag10} also find that the effective diffusion coefficients in the stratosphere, 
where convectively excited gravity waves are responsible for atmospheric mixing, also increase with increasing 
\teff\ for \teff\ $\le$ 1500 K and \teff\ $\ge$ 2000 K, but the behavior at intermediate 1500 $<$ \teff\ $<$ 2000 K 
becomes more complicated due to the effects of clouds, which alter atmospheric stability.  At the base of the 
stratosphere in the \citet{freytag10} models, the effective diffusion coefficient goes through a minimum.  
The \kzz\ profiles are also sensitive to gravity and the overall static stability in the atmosphere.  Without 
running realistic dynamical models for the planets in question, we cannot reliably estimate \kzz\ profiles 
\textit{a priori}, and we caution that our empirical profiles may have different magnitudes or functional forms 
than those of the real young-Jupiter atmospheres.  In particular, our profiles do not have the very weak \kzz\ 
minimum that might be expected at the base of the stratosphere on young Jupiters.  Because this minimum \kzz\ 
results in maximum column abundances for photochemical species produced at high altitudes \citep[e.g.,][]{bezard02}, 
our convention may cause us to underestimate the abundances of photochemical products, but not as severely as if 
we assumed that \kzz\ were constant throughout the atmosphere. 

\begin{figure}
\includegraphics[scale=0.5]{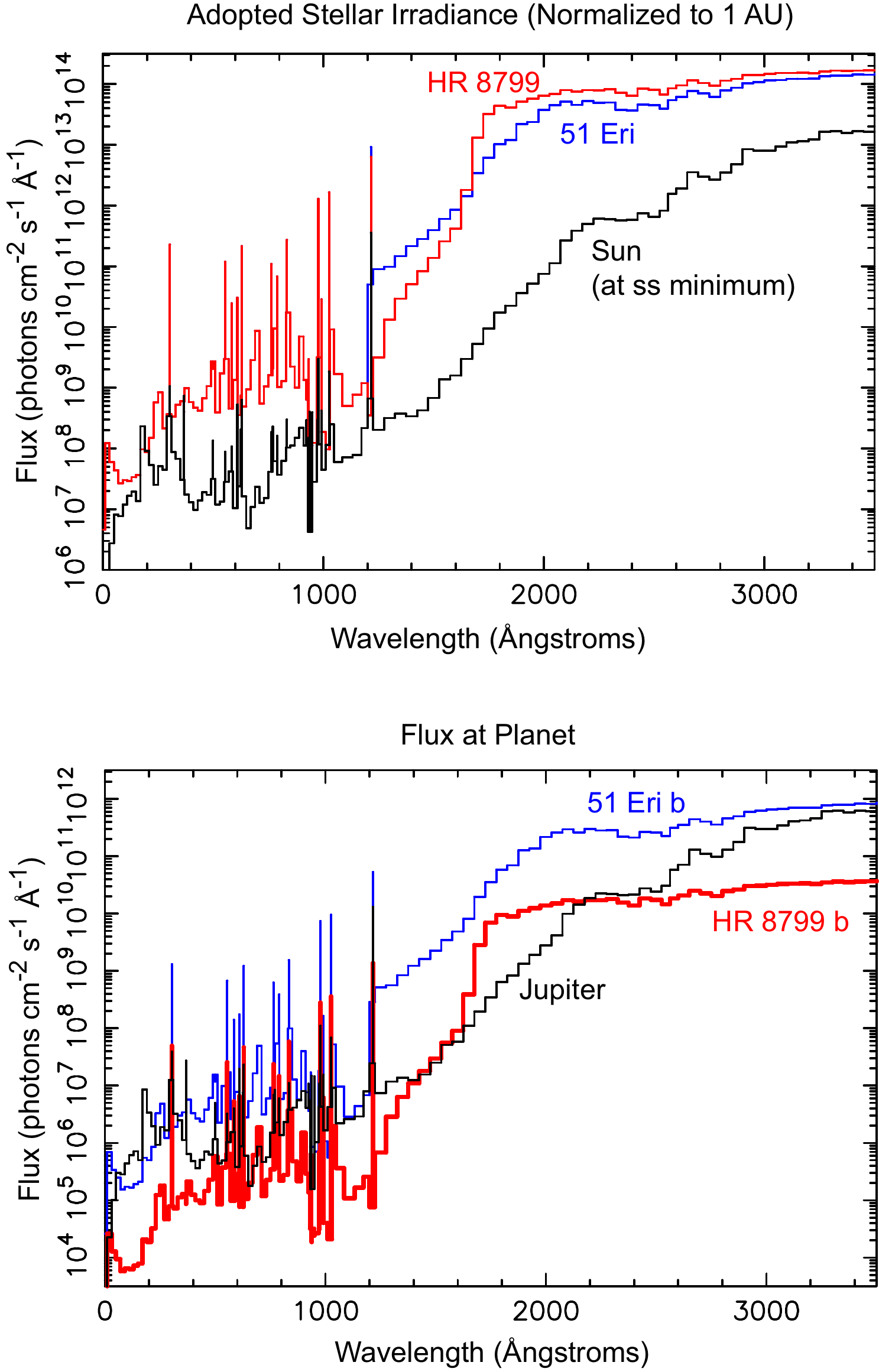}
\caption{The ultraviolet stellar irradiance adopted in the models: (Top) The irradiance of 51 Eri (blue) and 
HR 8799 (red) as received at 1 AU, in comparison with that the Sun (black); (Bottom) the irradiance at the 
top of the planet's atmosphere for 51 Eri b (blue) and HR 8799 b (red) in comparison with Jupiter (black).  
Note from the top panel that both 51 Eri and HR 8799 are brighter than the Sun in the ultraviolet, but 
51 Eri b and HR 8799 b are farther away from their host stars than Jupiter, so in terms of the H Lyman alpha 
flux received, which drives much of the interesting photochemistry, Jupiter receives a flux intermediate between 
51 Eri b and HR 8799 b (bottom panel).
A color version of this figure is available in the online journal.\label{figflux}}
\end{figure}

The photochemical model results also depend on the host star's ultraviolet flux and spectral energy distribution 
\citep[e.g.,][]{venot13,miguel14,miguel15}.  
For our specific exoplanet models, both 51 Eri (spectral type F0) and HR 8799 (spectral type A5) are expected 
to be brighter than the Sun at UV wavelengths (see Fig.~\ref{figflux}).  However, the only direct ultraviolet 
spectral observations we could find for either star are derived from International Ultraviolet Explorer (IUE) satellite 
observations of 51 Eri in the MAST archive (http://{\-}archive.{\-}stsci.{\-}edu).  Therefore, except for these 
IUE observations, our assumed stellar spectra are assembled from a variety of theoretical sources.  For wavelengths 
greater than 1979 \AA, the 51 Eri spectrum is taken from the \citet{heap11} NextGen model for 51 Eri (HD 29391); 
for wavelengths between 1200 and 1978.72 \AA\ --- except right at H Lyman $\alpha$ --- we use IUE observations of 
51 Eri from the MAST IUE archive; for wavelengths less than $\sim$1150 \AA, we adopt the theoretical spectrum of 
HR 8799 (as the closest analog star) from the \citet{sanzforcada11} X-exoplanets archive; and for Lyman $\alpha$ 
at 1215.7 \AA, we adopt the reconstructed intrinsic H Lyman alpha flux for 51 Eri from \citet{landsman93}.  
The HR 8799 spectrum is a composite of several theoretical models.  At wavelengths less than 1150 \AA\ and in the 
wavelength bin at 1190 \AA, the HR 8799 spectrum is from the aforementioned \citet{sanzforcada11} model of HR 8799; 
at wavelengths greater than 1150 \AA\ --- except for the wavelength bins at 1190 and 1215.7 \AA\ --- we use 
a \citet{castelli04} model with assumed parameters of $T_{\small\textsl{eff}}$ = 7500 K, log($g$) 
= 4.5 (cgs), log[Fe/H] = $-0.5$, radius = 1.44$R_{\odot}$; and for 1215.7 \AA, we estimate the flux as the 
average of four stars ($\kappa^2$ Tau [A7V], HR 1507 [F0V], 30 LMi [F0V], $\alpha$ Hyi [F0V]) from the 
\citet{landsman93} database of reconstructed intrinsic H Lyman alpha fluxes, after scaling appropriately for 
stellar distance.  For the spectral irradiance of the Sun shown in Fig.~\ref{figflux}, we adopt the solar-cycle minimum 
spectrum of \citet{woods02}.  

Note from Fig.~\ref{figflux} that 51 Eri and HR 8799 are intrinsically brighter than the Sun in 
the ultraviolet.  Despite the great orbital distances of the HR 8799  planets (b at $\sim$68 AU, c at 
$\sim$43 AU, d at $\sim$27 AU; cf. \citealt{marois08} \& \citealt{maire15}) and 51 Eri b (14 AU 
according to \citealt{derosa15}, although we used 13.2 AU for the calculations based on the earlier 
report by \citealt{macintosh15}), these planets --- like the giant planets within our own solar system --- 
receive sufficient ultraviolet flux that photochemistry should be effective.  In fact, 51 Eri b receives 
a greater H Lyman alpha flux than any of our solar-system giant planets, including Jupiter (see Fig.~\ref{figflux}), 
while the most distant HR 8799 b receives a greater H Ly $\alpha$ flux than either Uranus or Neptune, which both 
have rich stratospheric hydrocarbon photochemistry \citep{summers89,romani93,moses95c,dobrijevic10a,orton14chem}.  
Indeed, the first investigation into the photochemistry of 51 Eri b \citep{zahnle16} suggests that photochemical 
production of complex hydrocarbons and sulfur species will be important on this young Jupiter and may lead to the 
formation of sulfur and hydrocarbon hazes.

\section{Results\label{sectresults}}

Results from our thermo/photochemical kinetics and transport model are presented below.  We first 
discuss the results for generic directly-imaged planets, including trends as a function of $T_{\small\textsl{eff}}$, 
log($g$), $K_{\footnotesize\textsl{deep}}$, and distance from the host star \citep[see also][]{zahnle14}.  
The relevant disequilibrium chemistry that could potentially affect the spectral appearance of young Jupiters
is described.  Then, we present specific models for HR 8799 b and 51 Eri b and compare to observations.
Note that the model abundance profiles for both the generic and specific planets discussed below are included in the 
journal supplementary material.

\begin{figure*}[htb]
\begin{center}
\includegraphics[scale=0.8]{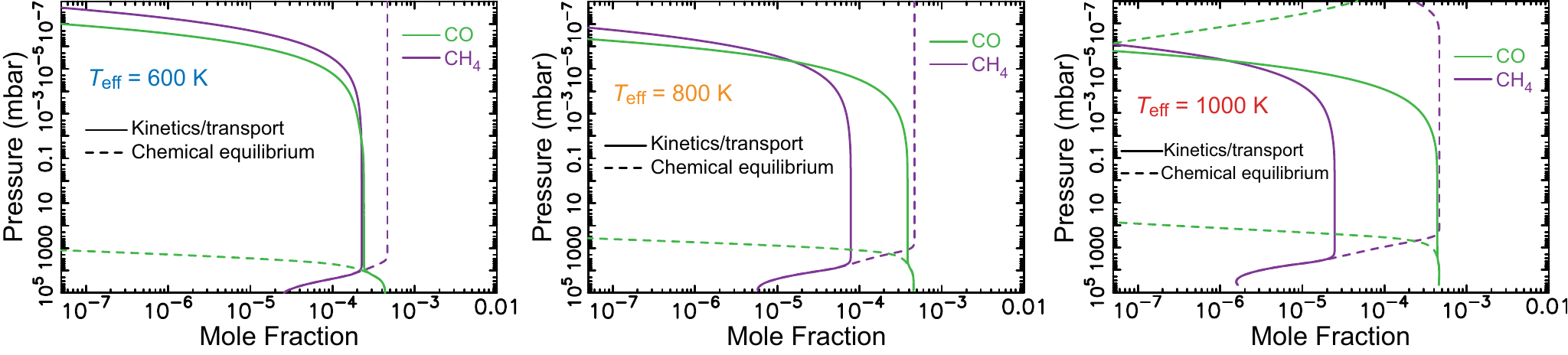}
\caption{The vertical  mixing-ratio profiles of CH$_4$ (purple) and CO (green) for planets with a surface gravity 
log($g$) = 4 (cgs), a moderate eddy mixing $K_{\footnotesize\textsl{deep}}$ = 10$^{7}$ cm$^{2}$ s$^{-1}$, and 
$T_{\footnotesize\textsl{eff}}$ = 600 K (Left), 800 K (Middle), 1000 K (Right).  Results for chemical equilibrium 
are shown with dashed lines, and results from our thermo/photochemical kinetics and transport model are shown as 
solid lines.  Note that CH$_4$ dominates in the observable portion of the atmosphere in chemical equilibrium, 
whereas CO dominates in the disequilibrium models.  The CH$_4$/CO ratio is strongly dependent on temperature for 
both types of chemistry, with a higher CH$_4$/CO ratio being favored for cooler planets.
A color version of this figure is available in the online journal.\label{figch4covsteff}}
\end{center}
\end{figure*}

\subsection{Generic Directly Imaged Planets: Chemistry\label{sectgeneric}}

For our ``generic'' young Jupiters, we generate a suite of models for nine different effective 
temperatures ($T_{\footnotesize\textsl{eff}}$ ranging from 600 K to 1400 K, at 100-K intervals), seven 
different eddy diffusion coefficient profiles (see Fig.~\ref{figeddy}), and two 
different surface gravities ($g$ = 10$^{3.5}$ and 10$^{4}$ cm s$^{-2}$).  The thermal profiles of these 
models are shown in Fig.~\ref{figtemp}.  Note from Fig.~\ref{figtemp} that all the 
models have deep atmospheres that lie within the CO stability field, whereas all but the hottest 
models switch over to the CH$_4$ stability field in the upper atmosphere.  Therefore, if 
the atmosphere were to remain in chemical equilibrium, CH$_4$ would be the dominant carbon 
constituent at ``photospheric'' pressures in the 10$^3$--0.1 mbar range for most of these planets, 
and methane absorption would be prominent in the near-infrared emission spectra.  However, CO 
$\leftrightarrows$ CH$_4$ chemical equilibrium cannot be maintained at temperatures $\lta$ 1300 K 
for any reasonable assumption about the eddy diffusion coefficient profile \citep[e.g.,][]{visscher11}, 
and quenching will occur in the deep, convective regions of these planets.  For all the thermal profiles 
investigated, the CO-CH$_4$ quench point occurs within the CO stability field, and the quenched 
abundance of CO will be greater than that of CH$_4$.

The dominant kinetic reaction scheme converting CO to CH$_4$ near the quench point in our models is
\begin{eqnarray}
\H \, + \, \CO \, + \, \M \, & \rightarrow & \, \HCO \, + \, \M \nonumber \\
\Htwo \, + \, \HCO \, & \rightarrow & \, \HtwoCO \, + \, \H \nonumber \\
\H \, + \, \HtwoCO \, + \, \M \, & \rightarrow & \, \CHtwoOH \, + \, \M \nonumber \\
\Htwo \, + \, \CHtwoOH \, & \rightarrow & \, \CHthreeOH \, + \, \H \nonumber \\
\CHthreeOH \, + \, \M \, & \rightarrow & \, \CHthree \, + \, \OH \, + \, \M \nonumber \\
\Htwo \, + \, \CHthree \, & \rightarrow & \, \CHfour \, + \, \H \nonumber \\
\Htwo \, + \, \OH \, & \rightarrow & \, \HtwoO \, + \, \H \nonumber \\
2\, \H \, + \, \M \, & \rightarrow & \, \Htwo \, + \, \M \nonumber \\
\noalign{\vglue -10pt}
\multispan3\hrulefill \nonumber \cr
\Net \ \ \CO \, + \, 3\, \Htwo \, & \rightarrow & \, \CHfour \, + \, \HtwoO  , \\
\end{eqnarray}
with M representing any third atmospheric molecule or atom.  This scheme is identical to the dominant scheme (15) 
that \citet{visscher11} propose is controlling the conversion of CO into CH$_4$ on brown dwarfs and is just
the reverse of the scheme (3) that \citet{moses11} propose is controlling $\CHfour$ $\rightarrow$ CO quenching on hot 
Jupiters.  The rate-limiting step 
in the above scheme is the reaction $\CHthreeOH$ + M $\rightarrow$ $\CHthree$ + OH + M, where the rate coefficient is 
derived from the reverse reaction from \citet{jasper07}.  Our chemical model differs from some others in the literature 
\citep[e.g.,][]{venot12,zahnle14} in that we adopt a slower rate coefficient for H + $\CHthreeOH$ $\rightarrow$ $\HtwoO$ + 
$\CHthree$ based on the \textit{ab initio} transition-state theory calculations of \citet{moses11} \& \citet{lendvay97}, and 
the discussion of relevant experimental data in \citet{norton90}.  However, the rate coefficient for this reaction 
adopted by \citet{zahnle14} and \citet{zahnle16} is slow enough that $\CHthreeOH$ + M $\rightarrow$ $\CHthree$ + OH + M 
is usually faster, and hence their quench results are not greatly different from those described here.
In any case, quenching is very effective in all the generic young-Jupiter models we investigated, and CO replaces 
CH$_4$ as the dominant carbon species in the photospheres of these planets.

\begin{figure*}[b]
\begin{center}
\includegraphics[scale=0.8]{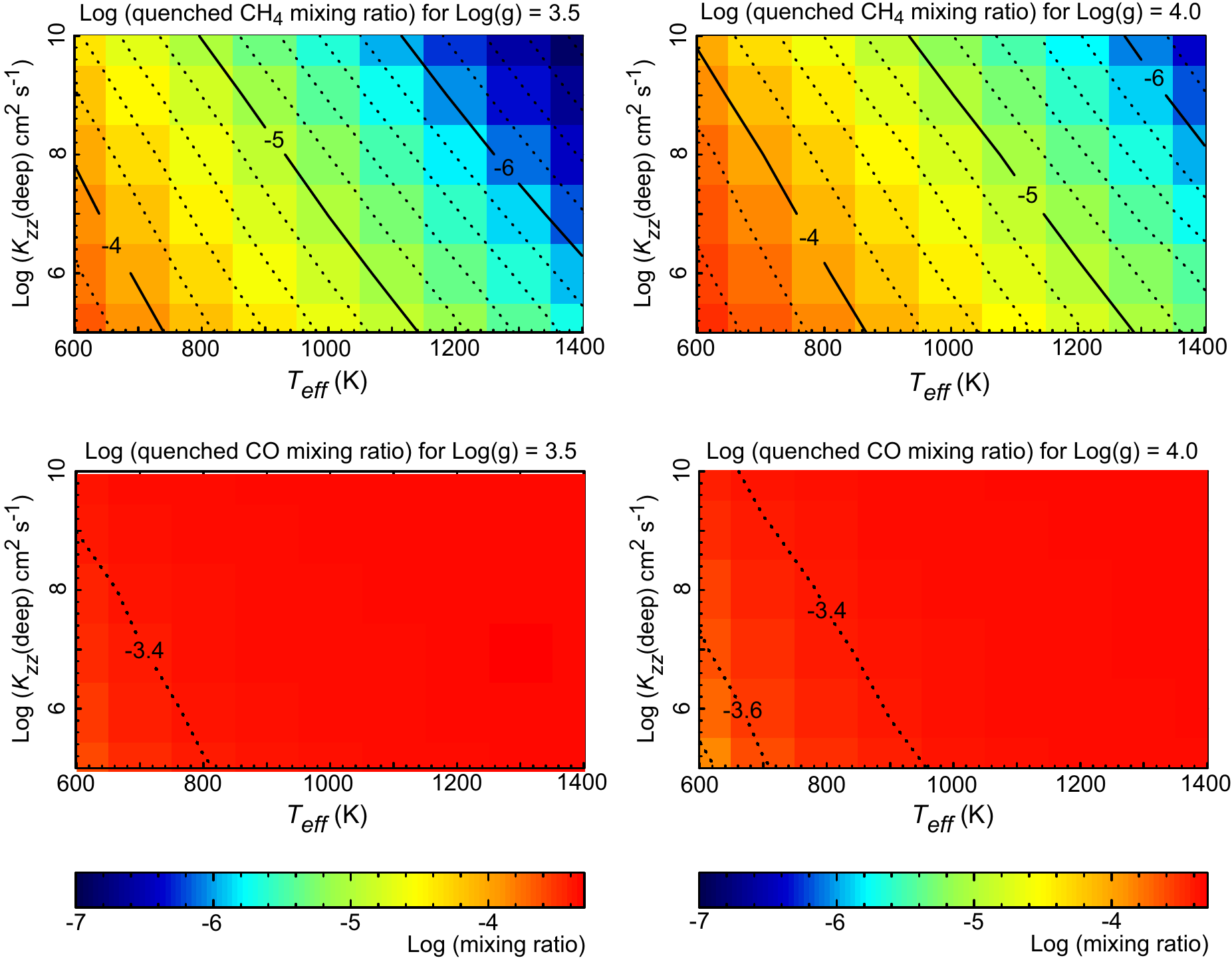}
\caption{Quenched mixing ratios of CH$_4$ (top) and CO (bottom) for models with surface gravities of $g$ = 10$^{3.5}$ 
(left) and 10$^{4}$ cm s$^{-2}$ (right) as a function of \teff\ and \kdeep.  High CH$_4$ abundances and low CO abundances 
are favored by small \teff, small \kdeep, and large $g$, although the CO abundance is relatively insensitive to these factors 
over the range of models investigated.
A color version of this figure is available in the online journal.\label{figgridall}}
\end{center}
\end{figure*}

\subsubsection{$\CO$-$\CHfour$ quenching as a function of \teff\ and \kzz}

Figure \ref{figch4covsteff} shows how the methane and carbon monoxide abundance vary with the planet's effective 
temperature (for \teff\ = 600, 800, 1000 K), for both the assumption of chemical equilibrium and from 
our thermo/photochemical kinetics and transport modeling, for \kdeep\ = 10$^7$ cm$^2$ s$^{-1}$ and 
log($g$) = 4 (cgs).
Figure~\ref{figch4covsteff} emphasizes just how significantly thermochemical equilibrium fails in its predictions 
for the composition of directly imaged planets, underpredicting the CO abundance by many orders of magnitude, and 
overpredicting the CH$_4$ abundance.  The CO-CH$_4$ quench point is discernible in the plot --- it is the pressure 
at which the CH$_4$ and CO mixing ratios stop following the equilibrium profiles and become constant with altitude.  
For the $T_{\footnotesize\textsl{eff}}$ = 600 K planet, the quench point is near the CO = CH$_4$ equal-abundance curve 
shown in Fig.~\ref{figtemp}, and carbon monoxide and methane quench at nearly equal abundances.  Warmer planets have 
quench points more solidly within the CO stability field, and so the CO abundance then exceeds that of methane at high 
altitudes.  The quenched CH$_4$ abundance depends strongly on \teff, decreasing with increasing \teff, when other 
factors like \kzz\ and $g$ are kept identical.  The depletion in both the CO and CH$_4$ mixing ratios at high altitudes 
in the disequilibrium models in Fig.~\ref{figch4covsteff} is due to molecular diffusion, which is dependent on temperature.  
Planets with a higher \teff\ have warmer upper atmospheres, causing molecular diffusion to take over at deeper 
levels.  Therefore, warmer planets have homopause levels at higher pressures (lower altitudes), all other things 
being equal.

The quenched species abundances also depend strongly on \kdeep\ and on surface gravity.  Figure \ref{figgridall} 
illustrates this relationship for CO (top row) and CH$_4$ (bottom row) for a suite of generic young 
Jupiter models, with the lower-gravity (log($g$) = 3.5) case being plotted in the left column and the 
higher-gravity case (log($g$) = 4.0) in the right column.  Note from Fig.~\ref{figgridall} that the quenched 
CH$_4$ abundance is highly sensitive to both \teff\ and \kdeep, and is greatest for low temperatures and weak 
deep vertical mixing.  Higher-gravity planets with the same \teff\ are cooler at any particular pressure level, so higher 
$g$ favors increased CH$_4$ abundance, all other  factors being equal.  In contrast, high $g$, low \teff, and low \kdeep\ 
favor \textsl{smaller} quenched CO abundances.  Note, however, the nearly constant quenched CO mixing ratio over a large 
swath of parameter space in Fig.~\ref{figgridall} for these two relatively low surface gravities.  The quenched CO mixing 
ratio is less sensitive than CH$_4$ to variations in 
\teff, \kdeep, and $g$ in this range because CO is dominant at the quench point, and 
the equilibrium CO mixing ratio is more constant with height through the quench region, whereas the equilibrium CH$_4$ 
mixing-ratio profile in this region has a significant vertical gradient.  

This is an important point.  Disequilibrium chemistry 
from transport-induced quenching will cause CO --- not CH$_4$ --- to dominate in the photospheres of virtually all directly imaged young 
planetary-mass (and planetary-gravity) companions, despite the equilibrium predictions for the predominance of $\CHfour$; in 
addition, the CO abundance should be similar for directly imaged planets with the same metallicity.  Spectral signatures of 
CO should therefore be common for young Jupiters, and derived CO abundances can help constrain the planet's metallicity.  
Note that this conclusion changes for higher-gravity ($g$ $\gta$ 10$^5$ cm s$^{-2}$) T dwarfs in this temperature range 
\citep{hubeny07,zahnle14}, where CH$_4$ can dominate and CO is the minor species.

\begin{figure*}
\begin{center}
\includegraphics[scale=0.8]{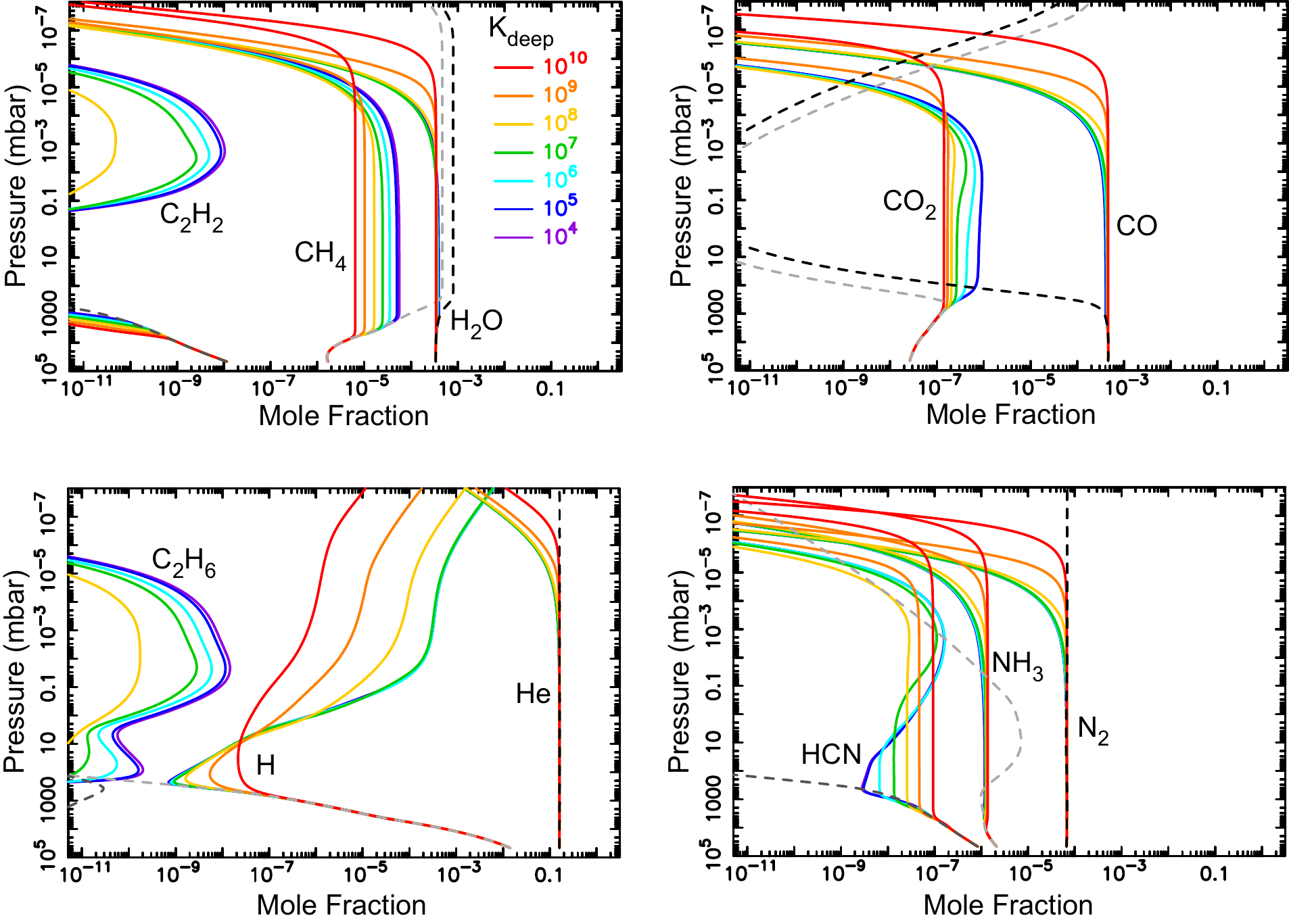}
\caption{Vertical profiles of several important species in our thermo/photochemical kinetics and transport models 
(solid colored lines) and in chemical equilibrium (dashed gray and black lines) for a planet with \teff\ = 1000 K 
and $g$ = 10$^{4}$ cm s$^{-2}$, at a distance of 68 AU from a star with properties like HR 8799 (Fig.~\ref{figflux}), 
as a function of \kdeep\ (see the legend in the top left panel, and the \kzz\ profiles shown in Fig.~\ref{figeddy}).  
Note that the atmosphere is far out of equilibrium for all the eddy diffusion coefficient profiles considered.  The 
quenched CH$_4$ mixing ratio increases with decreasing \kdeep.  The mixing ratios of methane photochemical 
products such as C$_2$H$_2$, C$_2$H$_6$, and H also increase with decreasing \kdeep.  Water quenches at the same time 
as CO and CH$_4$, remaining in disequilibrium in the photosphere.  Species like HCN and CO$_2$ are affected both by
photochemistry and by quenching of the major carbon, oxygen, and nitrogen species.
A color version of this figure is available in the online journal.\label{figmixvseddy}}
\end{center}
\end{figure*}

\subsubsection{Sensitivity of disequilibrium chemistry to \kzz}

Figure \ref{figmixvseddy} 
illustrates how the abundances of several constituents change with the different eddy diffusion coefficient profiles 
shown in Fig.~\ref{figeddy}, for a model with \teff\ = 1000 K, log($g$) = 4.0, and an orbital distance of 68 AU from 
a star with the properties of HR 8799.  As the eddy diffusion coefficient at depth, \kdeep, is increased, vertical 
transport begins to dominate at greater and greater depths over the chemical kinetic reactions that act to maintain 
equilibrium.  Smaller \kdeep\ values lead to mixing ratio profiles that follow the equilibrium profiles to higher 
altitudes before quenching occurs.  The quenched methane abundance therefore increases with decreasing \kdeep, and 
species that are produced through the photochemical destruction of methane, like C$_2$H$_2$ and C$_2$H$_6$, also have 
mixing ratios that increase with decreasing \kdeep.  Conversely, the quenched CO abundance decreases with decreasing 
\kdeep, but because the chemical equilibrium abundance of CO is only slightly decreasing with altitude over the range 
of quench points for the different \kdeep\ values investigated, the quenched CO mixing ratio is relatively insensitive 
to \kdeep.  

Water quenches via reaction scheme (1) above at the same point as that of CO and CH$_4$.  Since the equilibrium mixing ratio 
for H$_2$O is increasing with increasing altitude very slightly over the pressure range of the quench points, the quenched 
H$_2$O abundance very slightly increases with decreasing \kdeep.  Water is a key opacity source in young Jupiters that 
affects how efficiently heat is lost from the planet, so it is important to keep in mind that the resulting quenched 
water mixing ratio on directly imaged planets can be a factor of a few below that of chemical-equilibrium predictions 
in the photosphere.  This quenching of H$_2$O becomes more important for higher \teff, larger \kdeep, and lower surface 
gravities.  Quenching of water should therefore be considered in models that calculate the thermal evolution of brown dwarfs 
and directly imaged planets, particularly for young, small, hot objects.

The NH$_3$-N$_2$ quench point is deeper than that of CO-CH$_4$-H$_2$O.  For all the planets considered, this major 
nitrogen-species quench point is well within the N$_2$-dominated regime, so N$_2$ dominates in the photosphere, 
and NH$_3$ is less abundant.  The equilibrium profiles are not strongly sloped in the quench region, so the 
quenched abundances of NH$_3$ --- and N$_2$ in particular --- are not very sensitive to \kdeep\ (see Fig.~\ref{figmixvseddy}).
The dominant quenching scheme for N$_2$ $\rightarrow$ NH$_3$ in our generic young-Jupiter models is
\begin{eqnarray}
\H \, + \, \Ntwo \, + \, \M \, & \rightarrow & \, \NtwoH \, + \, \M \nonumber \\
\Htwo \, + \, \NtwoH \, & \rightarrow & \, \NtwoHtwo \, + \, \H \nonumber \\
\H \, + \, \NtwoHtwo \, & \rightarrow & \, \NH \, + \, \NHtwo \nonumber \\
\Htwo \, + \, \NH \, & \rightarrow & \, \NHtwo \, + \, \H \nonumber \\
2\, ( \, \Htwo \, + \, \NHtwo \, & \rightarrow & \, \NHthree \, + \, \H \, ) \nonumber \\
2\, \H \, + \, \M \, & \rightarrow & \, \Htwo \, + \, \M \nonumber \\
\noalign{\vglue -10pt}
\multispan3\hrulefill \nonumber \cr
\Net \ \ \Ntwo \, + \, 3\, \Htwo \, & \rightarrow & \, 2\, \NHthree , \\
\end{eqnarray}
which is simply the reverse of reaction scheme (5) for $\NHthree$ $\rightarrow$ $\Ntwo$ quenching discussed in \citet{moses11}.  
The rate-limiting step in the 
above scheme is the reaction H + $\NtwoHtwo$ $\rightarrow$ NH + $\NHtwo$, where the rate coefficient derives from 
the reverse reaction, as determined by \citet{klippenstein09}.

Constituents such as HCN and CO$_2$ are affected both by photochemistry and by quenching of the dominant 
carbon, nitrogen, and oxygen carriers (H$_2$O, CO, CH$_4$, NH$_3$, and N$_2$) and thus  exhibit complicated vertical 
profiles in Fig.~\ref{figmixvseddy}.  For large values of \kzz$(z)$, transport controls the HCN and CO$_2$ 
profiles throughout the atmospheric column.  The quenched abundance of HCN increases with increasing \kdeep\ 
because the equilibrium profile decreases with height within the quench region.  Conversely, the quenched abundance
of CO$_2$ decreases with increasing \kdeep\ because the equilibrium profile increases with height near the 
quench point; moreover, the photochemically produced CO$_2$ takes longer to diffuse downward when the stratospheric 
\kzz\ is smaller, so a larger column abundance can build up.  In fact, at higher altitudes with the smaller \kdeep\ 
models, photochemical production of HCN and CO$_2$ can dominate over transport from below, and the resulting mixing-ratio 
``bulges'' in the stratosphere represent the signatures of that photochemical production.  In general, the column-integrated 
CO$_2$ abundance increases with decreasing \kdeep, while that of HCN decreases with decreasing \kdeep.  However, this 
latter result also depends on the planet's thermal structure and incident ultraviolet flux.

Note that the sharp drop off in the species profiles at high altitudes in Fig.~\ref{figmixvseddy} is due to molecular 
diffusion.  Because the molecular diffusion coefficient profiles for this thermal structure cross the \kzz\ profiles 
at relatively high altitudes where the \kzz\ profiles have already transitioned to the $P^{0.5}$ sloped region, the
homopause levels for most of the models for any particular species are the same for the different \kzz\ models.  However,
the CH$_4$ homopause level is at $\sim$3$\scinot-7.$ mbar for the sloped \kzz\ case, and Fig.~\ref{figeddy} shows that 
the \kdeep\ = 10$^{10}$ cm$^2$ s$^{-1}$ \kzz\ profile does not reached the sloped \kzz\ portion until pressures less 
than a few $\times$ 10$^{-8}$ mbar.  Therefore, the CH$_4$ molecular diffusion coefficient crosses the \kdeep\ = 
10$^{10}$ cm$^2$ s$^{-1}$ \kzz\ profile at a higher altitude (lower pressure) than the other models, leading to a 
higher-altitude homopause and CH$_4$ being carried to higher altitudes in that model than the others.  Similarly, the 
H$_2$O, NH$_3$, CO, and N$_2$ molecular diffusion coefficients cross the sloped \kzz\ profile at pressures between 
where the \kdeep\ = 10$^{8}$ and 10$^{9}$ cm$^2$ s$^{-1}$ models transition to the sloped case, so both the \kzz\ = 
10$^{9}$ and 10$^{10}$ cm$^2$ s$^{-1}$ cases have higher-altitude H$_2$O, NH$_3$, CO, and N$_2$ homopauses than the 
other models.

\begin{figure}[htb]
\includegraphics[scale=0.8]{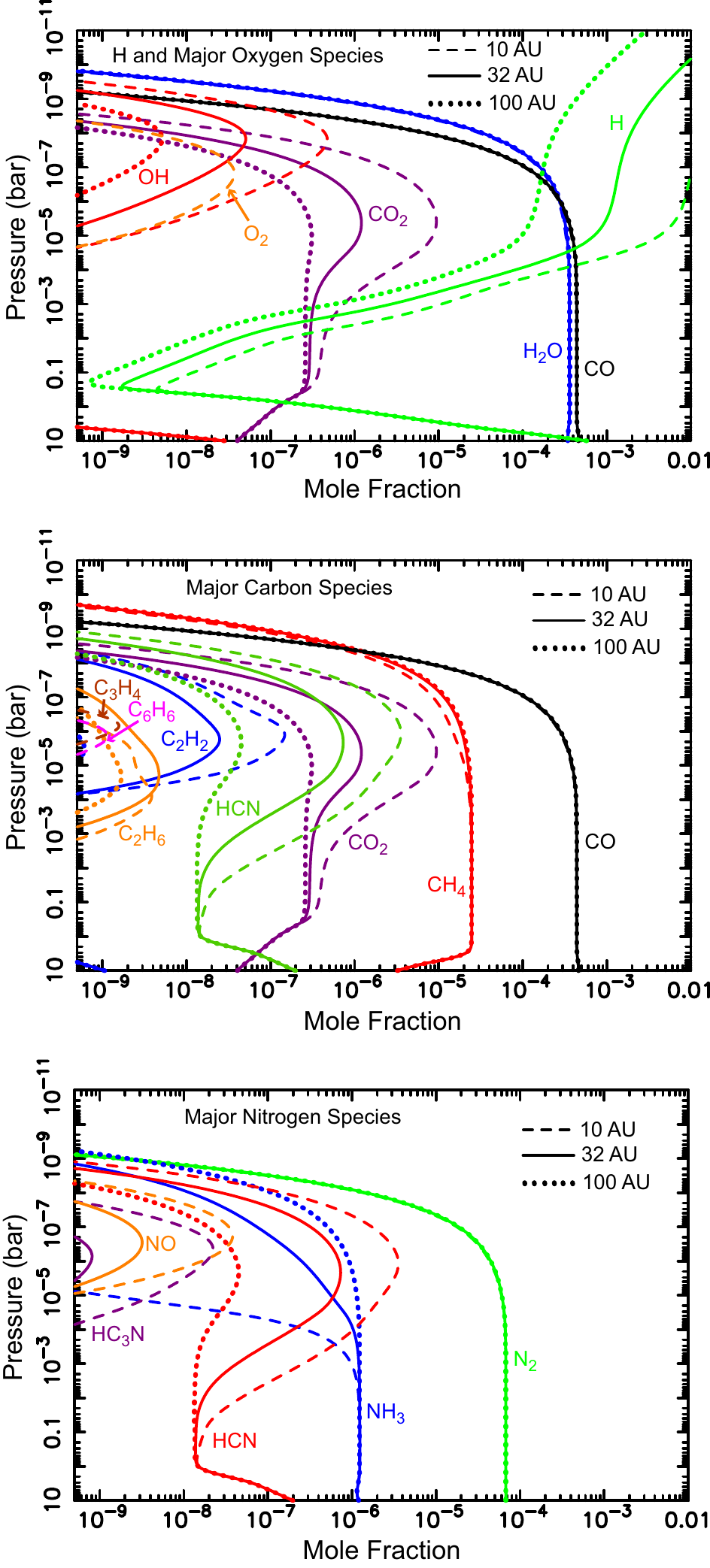}
\caption{The vertical mixing-ratio profiles of several atmospheric species as a function of orbital distance for 
a planet with \teff\ = 1000 K, $g$ = 10$^4$ cm s$^{-2}$, and \kdeep\ = 10$^7$ cm$^2$ s$^{-1}$, that is being 
irradiated by an HR 8799-like star at a distance of 10 AU (dashed lines), 32 AU (solid lines), and 100 AU (dotted 
lines).  The greater UV flux received by the closest-in planet leads to increased destruction of photochemically 
active ``parent'' molecules such as CH$_4$, NH$_3$, H$_2$O, CO, and N$_2$, and increased production of 
photochemical ``daughter'' products such as HCN, CO$_2$, complex hydrocarbons, complex nitriles, and atomic species 
and small radicals.
A color version of this figure is available in the online journal.\label{figdistance}}
\end{figure}

\subsubsection{Sensitivity of disequilibrium chemistry to orbital distance}

Figure \ref{figdistance} illustrates how the disequilibrium composition changes as a function of distance from the 
host star, for planets with \teff\ = 1000 K, log($g$) = 4.0 (cgs), \kdeep\ = 10$^7$ cm$^2$ s$^{-1}$, orbiting 
at 10, 32, and 100 AU from a star with the properties of HR 8799.  Because the strong interior heat 
dominates the energy transport on these young planets, the thermal structures are virtually identical in these 
cases, so the main differences in the models are due to the incoming ultraviolet flux. The closer a planet 
is to its star, the greater the UV irradiation received, leading to greater 
destruction rates of key molecules such as CH$_4$, NH$_3$, H$_2$O, CO, and N$_2$.  That in turn leads to greater 
production rates of photochemical products such as HCN, CO$_2$, C$_2$H$_2$, C$_2$H$_6$, complex hydrocarbons such as 
methylacetylene (an isomer of C$_3$H$_4$) and benzene (an isomer of C$_6$H$_6$), complex nitriles such as HC$_3$N, 
small oxygen-bearing species such as NO and O$_2$, and small radicals and atoms such as C, N. O, OH, NH$_2$, and CH$_3$.

The dominant photochemical product on young Jupiters is atomic hydrogen.  The atomic H is derived largely from 
water photolysis (producing OH + H), and the subsequent reaction of OH + H$_2$ $\rightarrow$ H$_2$O + H --- a two-step 
process that 
catalytically destroys H$_2$ to produce two H atoms.  In this regard, young Jupiters have more in common with close-in 
transiting giant planets \citep[e.g.,][]{liang03} than our solar-system giant planets, and the copious amount of atomic 
H produced from this photochemistry (see Fig.~\ref{figdistance}) affects much of the subsequent stratospheric chemistry 
on young Jupiters.  

Another key photochemical product is CO$_2$.  Carbon dioxide is 
produced overwhelmingly from the reaction OH + CO $\rightarrow$ CO$_2$ + H, with the OH deriving from water photolysis.  
If the stratosphere is relatively warm, as in the example shown in Fig.~\ref{figdistance} (with a 1 $\mu$bar temperature 
of 377 K), the OH + H$_2$ $\rightarrow$ H$_2$O + H reaction occurs at a much faster rate than OH + CO $\rightarrow$ 
CO$_2$ + H, but the latter reaction provides a slow but steady stream of oxygen away from water and CO into CO$_2$.  
Loss of CO$_2$ occurs through the reverse of the main production reaction (i.e., H + CO$_2$ $\rightarrow$ CO + OH), 
provided that the upper-atmospheric temperature is warm enough to overcome the substantial energy barrier for this 
reaction, as well as through photolysis, through reaction of atomic N to produce NO + CO, and through reaction of CH 
to produce HCO + CO.  Note that all the main loss processes for CO$_2$ end up recycling the CO.  For our generic 
young Jupiter models, the column-integrated CO$_2$ production rate exceeds the loss rate, and the photochemically 
produced CO$_2$ diffuses down through the atmosphere until it reaches higher-temperature regions
where it can once again reach a chemical balance with CO and H$_2$O.  The greater the incident ultraviolet flux, 
the greater the net photochemical production rate of CO$_2$ (see Fig.~\ref{figdistance}).

Molecular oxygen becomes a notable high-altitude photochemical product on more highly-irradiated young 
Jupiters.  It is produced as a byproduct of the water photochemistry, where photolysis of H$_2$O produces 
OH + H and O + 2H, and the OH and O react to form O$_2$ + H.  The O$_2$ is lost through photolysis (which 
primarily leads back to H$_2$O eventually) and through reactions with atomic carbon (which leads to CO).

Some of the CH$_4$ in the upper atmospheres of young Jupiters will be oxidized to produce CO and eventually 
CO$_2$.  In our generic young Jupiter models, this process occurs through schemes such as:
\begin{eqnarray}
\HtwoO \, + \, h\nu \, & \rightarrow & \, 2\, \H \, + \, \O \nonumber \\
\H \, + \, \CHfour \, & \rightarrow & \, \CHthree \, + \, \Htwo \nonumber \\
\O \, + \, \CHthree \, & \rightarrow & \, \HtwoCO \, + \, \H \nonumber \\
\HtwoCO \, + \, \H \, & \rightarrow & \, \HCO \, + \, \Htwo \nonumber \\
\HCO \, + \, \H \, & \rightarrow & \, \CO \, + \, \Htwo \nonumber \\
\noalign{\vglue -10pt}
\multispan3\hrulefill \nonumber \cr
\Net \ \ \CHfour \, + \, \Htwo \, & \rightarrow & \, \CO \, + \, 3\, \Htwo , \\
\end{eqnarray}
with $h\nu$ representing an ultraviolet photon.  Methane oxidation schemes such as the one above are more effective the 
higher the incident stellar ultraviolet flux.

As on the giant planets in our own solar system \citep[e.g.,][]{strobel83,atreya85,yung99,moses04,fouchet09}, the reduced 
hydrocarbon photochemistry in the atmospheres of young Jupiters will be efficacious and complex.  However, the overall column 
abundance of the hydrocarbon species produced by neutral photochemistry (as opposed to ion chemistry) on young Jupiters will 
typically be smaller than on our own giant planets, as a result of the greater stratospheric temperatures, greater
stratospheric water abundance, and different dominant 
and/or competing kinetic reactions, including methane recycling and oxidation.  The typically smaller CH$_4$ mixing ratio on young 
Jupiters (due to quenching) also contributes to the differences, as does a potentially larger stratospheric eddy \kzz\ coefficient 
(due to upwardly propagating atmospheric waves generated in the rapidly convecting deep atmospheres of young Jupiters), which allows 
the high-altitude hydrocarbon photochemical products to be transported more rapidly to the deeper, high-temperature regions, where 
they become unstable.  However, the larger stratospheric temperatures and resulting decreased stability of the complex hydrocarbons 
play a larger role.  

As an example, the column abundance of ethane (C$_2$H$_6$) above 100 mbar on Saturn \citep{moses15}, which is 
$\sim$10 AU from the Sun, is five orders of magnitude larger than that of the generic 10-AU young Jupiter shown in 
Fig.~\ref{figdistance}, despite the greater H Lyman alpha and overall UV flux received by the 10-AU generic young Jupiter around 
its brighter star.  The main source of the ethane is still the same on both planets --- the three-body reaction CH$_3$ + CH$_3$ + M 
$\rightarrow$ C$_2$H$_6$ + M --- but the CH$_3$ on the 10-AU young Jupiter goes back to recycle the CH$_4$  more than 99.9\% of the 
time, because the higher atmospheric temperatures lead to a more efficient reaction of CH$_3$ with H$_2$ to form CH$_4$ + H.  Still, 
the total stratospheric column production rate of C$_2$H$_6$ is larger on the 10-AU young Jupiter than on Saturn due to the brightness 
of the star and the larger UV flux; however, C$_2$H$_6$ is also more readily destroyed on the warmer young Jupiter through H + 
C$_2$H$_6$ $\rightarrow$ C$_2$H$_5$ + H$_2$, with a much larger percentage of the carbon ending up back in CH$_4$ rather than in 
C$_2$H$_x$ and other higher-order hydrocarbons.  On Saturn, the photochemically produced C$_2$H$_6$ is much more chemically stable 
in the colder stratosphere, so the net production rate minus loss rate is greater on Saturn than on the generic 10-AU young Jupiter.
It is also interesting to note that the direct photolysis of CH$_4$ on our warmer generic young Jupiters is less important to the 
production of complex hydrocarbons than the reaction of atomic H with CH$_4$ to form CH$_3$ + H$_2$, with the H deriving from H$_2$O 
photolysis (see discussion above).  

Acetylene (C$_2$H$_2$) is also an important photochemical product on our 10-AU generic young Jupiter shown in 
Fig.~\ref{figdistance} that is produced through reaction schemes such as the following that first go through 
C$_2$H$_6$ and C$_2$H$_4$:
\begin{eqnarray}
2\, ( \, \HtwoO \, + h\nu & \rightarrow & \, \OH \, + \, \H \, ) \nonumber \\
2\, ( \, \OH \, + \Htwo & \rightarrow & \, \HtwoO \, + \, \H \, ) \nonumber \\
2\, ( \, \H \, + \CHfour & \rightarrow & \, \CHthree \, + \, \Htwo \, ) \nonumber \\
\CHthree \, + \, \CHthree \, + \, \M \, & \rightarrow & \, \CtwoHsix \, + \, \M \nonumber \\
\H \, + \, \CtwoHsix \, & \rightarrow & \, \CtwoHfive \, + \, \Htwo \nonumber \\
\CtwoHfive \, + \, \M \, & \rightarrow & \, \CtwoHfour \, + \, \H \, + \, \M \nonumber \\
\H \, + \, \CtwoHfour  \, & \rightarrow & \, \CtwoHthree \, + \, \Htwo \nonumber \\
\H \, + \, \CtwoHthree \, & \rightarrow & \, \CtwoHtwo \, + \, \Htwo \nonumber \\
\noalign{\vglue -10pt}
\multispan3\hrulefill \nonumber \cr
\Net \ \ 2\, \CHfour \, & \rightarrow & \, \CtwoHtwo \, + \, 3\, \Htwo . \\
\end{eqnarray}
Acetylene is lost (a) through insertion reactions with atomic C and CH radicals 
to form C$_3$H$_2$ and C$_3$H$_3$, (b) through reactions with atomic H to form C$_2$H$_3$, with subsequent reactions leading to other 
C$_2$H$_x$ species and eventual methane recycling, and (c) by photolysis, which leads predominantly to recycling of the 
C$_2$H$_2$.  As on transiting hot Jupiters \citep{moses11}, the atomic carbon from loss process (a) here derives both from
photolysis of CO and from methane photodestruction to form CH$_3$, CH$_2$, and CH, which can react with H to eventually form C.

The relative efficiency of C$_3$H$_2$ and C$_3$H$_3$ production in some of our more highly irradiated young-Jupiter models (e.g, 
the 10-AU case) is interesting and suggests that complex carbon-rich species like PAHs could 
potentially form on some directly imaged planets, and might even lead to the condensation of organic hazes in these atmospheres, 
as enthusiastically advocated by \citet{zahnle09soot,zahnle16}.  However, in general, the efficiency of production of refractory 
organics from simple precursors like C$_2$H$_2$, C$_2$H$_6$, and C$_4$H$_2$ in an H$_2$-dominated atmosphere seems to have been 
overestimated by \citet{zahnle09soot}, \citet{miller-ricci12}, and \citet{morley13} --- their arguments 
would suggest that Jupiter, Saturn, and Neptune should be completely enshrouded in optically thick stratospheric hydrocarbon 
hazes, yet that is not the case.  Because of a lack of laboratory or theoretical kinetic information on reactions of C$_3$H$_2$ and 
C$_3$H$_3$ with other hydrocarbon radicals under relevant low-pressure, reducing conditions, the fate of these C$_3$H$_x$ species 
is not obvious \citep[see also][]{moses11,hebrard13}.  Three-body addition reactions of C$_3$H$_2$ and C$_3$H$_3$ with abundant ambient 
H atoms can lead to C$_3$H$_3$ and C$_3$H$_4$, respectively, and the C$_3$H$_3$ can react with CH$_3$ to form C$_4$H$_6$ 
\citep{fahr00,knyazev01} or self-react to form various C$_6$H$_6$ isomers \citep{atkinson99,fahr00}, but these three-body 
reactions are not particularly effective at low pressures.  Therefore, C$_3$H$_2$ and C$_3$H$_3$ build up to mixing ratios of a few 
$\times$ 10$^{-8}$ at high altitudes in our 10-AU young-Jupiter model.  The comparatively large abundance of C$_3$H$_2$ and 
C$_3$H$_3$ radicals here is likely an artifact of having insufficient knowledge of other possible loss mechanisms for these 
species, and we make a plea for future laboratory experiments or theoretical modeling to rectify this situation.

Benzene (C$_6$H$_6$) itself is produced in our models through C$_3$H$_3$--C$_3$H$_3$ recombination, which 
first goes through a linear C$_6$H$_6$ isomer before eventual production of benzene \citep{fahr00}.  The benzene mixing ratio 
reaches 1 ppb in our 10-AU model (see Fig.~\ref{figdistance}), but neither benzene nor any of the other relatively 
light hydrocarbons considered by our model become 
abundant enough to achieve saturation and condense.  Similarly, the coupled carbon-nitrogen photochemistry in our model leads 
to non-trivial amounts of complex nitriles such as HC$_3$N being produced (see Fig.~\ref{figdistance}), but again, these 
relatively light nitriles never reach saturation.  Our neutral chemistry alone does not lead to hazes on these planets.  
However, we know from Titan that organic hazes can readily form from ion chemistry in a N$_2$-dominated atmosphere 
\citep{waite07,vuitton07,imanaka07,horst12}, and the presence of $>$ 10 ppm N$_2$ in the upper atmospheres of young Jupiters 
may augment the production of refractory condensable hydrocarbons through Titan-like ion chemistry.  This possibility deserves 
further investigation, both experimentally and theoretically.

The dominant product of the coupled carbon-nitrogen photochemistry is HCN, which forms through hypothesized schemes such 
as the following:
\begin{eqnarray}
\Ntwo \, + h\nu & \rightarrow & \, 2\, \N  \nonumber \\
\HtwoO \, + h\nu & \rightarrow & \, \OH \, + \, \H \nonumber \\
\OH \, + \Htwo & \rightarrow & \, \HtwoO \, + \, \H \nonumber \\
2\, ( \, \H \, + \CHfour & \rightarrow & \, \CHthree \, + \, \Htwo \, ) \nonumber \\
2\, ( \, \N \, + \CHthree & \rightarrow & \, \HtwoCN \, + \, \H \, ) \nonumber \\
2\, ( \, \HtwoCN \, + \H & \rightarrow & \, \HCN \, + \, \Htwo \, ) \nonumber \\
\noalign{\vglue -10pt}
\multispan3\hrulefill \nonumber \cr
\Net \ \ \Ntwo \, + \, 2\, \CHfour \, & \rightarrow & \, 2\, \HCN \, + \, 3\, \Htwo . \\
\end{eqnarray}
Note that N$_2$, not NH$_3$, is the source of the nitrogen in this scheme, which is effective at high altitudes.  That is 
why the HCN abundance can exceed the NH$_3$ abundance at high altitudes in the 10-AU model shown in Fig.~\ref{figdistance}.  
However, NH$_3$ can also contribute to HCN formation through schemes such as the following that are more effective at 
lower stratospheric altitudes:
\begin{eqnarray}
\NHthree \, + h\nu & \rightarrow & \, \NHtwo \, + \, \H  \nonumber \\
2\, (\, \HtwoO \, + h\nu & \rightarrow & \, \OH \, + \, \H \, ) \nonumber \\
2\, (\, \OH \, + \Htwo & \rightarrow & \, \HtwoO \, + \, \H \, ) \nonumber \\
\H \, + \CHfour & \rightarrow & \, \CHthree \, + \, \Htwo \nonumber \\
\CHthree \, + \NHtwo \, + \, \M & \rightarrow & \, \CHthreeNHtwo \, + \, \M \, ) \nonumber \\
\CHthreeNHtwo \, + \H & \rightarrow & \, \CHtwoNHtwo \, + \, \Htwo \nonumber \\
\CHtwoNHtwo \, + \H & \rightarrow & \, \CHtwoNH \, + \, \Htwo \nonumber \\
\CHtwoNH \, + \H & \rightarrow & \, \HtwoCN \, + \, \Htwo \nonumber \\
\HtwoCN \, + \H & \rightarrow & \, \HCN \, + \, \Htwo \nonumber \\
\noalign{\vglue -10pt}
\multispan3\hrulefill \nonumber \cr
\Net \ \ \NHthree \, + \, \CHfour \, & \rightarrow & \, \HCN \, + \, 3\, \Htwo . \\
\end{eqnarray}
As shown in Fig.~\ref{figdistance}, the coupled nitrogen-carbon photochemistry is more efficient with a greater UV 
flux from the host star.

Molecular nitrogen is fairly stable on young Jupiters.  Photodissociation is only effective at wavelengths shorter than 
$\sim$1000 \AA, so N$_2$ can be shielded to some extent by the more abundant H$_2$, CO, and H$_2$O.   In addition, the atomic 
N produced from N$_2$ photolysis can go back to recycle the N$_2$, through reactions such as N + OH $\rightarrow$ NO + 
H, followed by N + NO $\rightarrow$ N$_2$ + O.  The production rate of NO through this process exceeds the 
loss rate, and NO appears as a minor high-altitude photochemical product on young Jupiters (Fig.~\ref{figdistance}),  
especially for higher UV irradiation levels.

Ammonia, on the other hand, is much less stable than N$_2$ because of weaker bonds, photolysis out to longer wavelengths 
($\lambda$ $\lta$ 2300 \AA), efficient reaction with atomic H, and relatively inefficient recycling.  The NH$_3$ photolysis 
products can end up in N$_2$ through reactions such as N + NH$_2$ $\rightarrow$ NNH + H, followed by NNH $\rightarrow$ N$_2$ + H, or 
by NH$_2$ + H $\rightarrow$ NH + H$_2$, followed by NH + H $\rightarrow$ N + H$_2$, and N + NO $\rightarrow$ N$_2$ + O.  
The nitrogen in the ammonia can also end up in HCN, through reaction pathways such as scheme (6) above.  As is apparent 
from Fig.~\ref{figdistance}, the NH$_3$ in the upper stratosphere of young Jupiters becomes more depleted the higher the 
incident UV flux.  

One other nitrogen-bearing photochemical product worth mentioning is HC$_3$N, which is produced in the model through 
reaction of atomic N with C$_3$H$_2$ and C$_3$H$_3$ \citep[e.g.,][]{millar91} --- speculative reactions that may not 
be as efficient if we had more information about additional loss processes for these C$_3$H$_x$ species --- and by 
CN + $\CtwoHtwo$ $\rightarrow$ HC$_3$N + H (with the CN from HCN photolysis), which at least has a more convincing 
pedigree \citep[e.g.,][]{sims93}.  Again, more HC$_3$N (and CH$_3$CN) are produced with higher incident UV fluxes.  
We have not included in the model reactions from the coupled photochemistry of C$_2$H$_2$ and NH$_3$, which can produce 
a host of complex organic molecules \citep[e.g.,][]{keane96,moses10}, due to a lack of published thermodynamic properties 
for these molecules.  However, heavier species such as acetaldazine, acetaldehyde hydrazone, and ethylamine may also form on 
young Jupiters due to this coupled chemistry, particularly on cooler, more highly UV irradiated planets.  Unlike on
our own solar-system gas giants, hydrazine (N$_2$H$_4$) is not a major product of the ammonia photochemistry in our 
young-Jupiter models because the NH$_2$ from ammonia photolysis preferentially reacts with the copious amounts of atomic H 
to produce NH, and eventually N and N$_2$, or with CH$_3$ to form CH$_3$NH$_2$ and eventually HCN.  On Jupiter and 
Saturn, the coupled ammonia-methane photochemistry is less efficient due to the lack of CH$_3$ present in the tropospheric 
region where NH$_3$ is photolyzed \citep[e.g.,][]{kaye83ch3nh2,moses10}.  However, the hydrazine abundance is very sensitive 
to temperature and increases significantly as \teff\ decreases.

\begin{figure}[htb]
\includegraphics[scale=0.9]{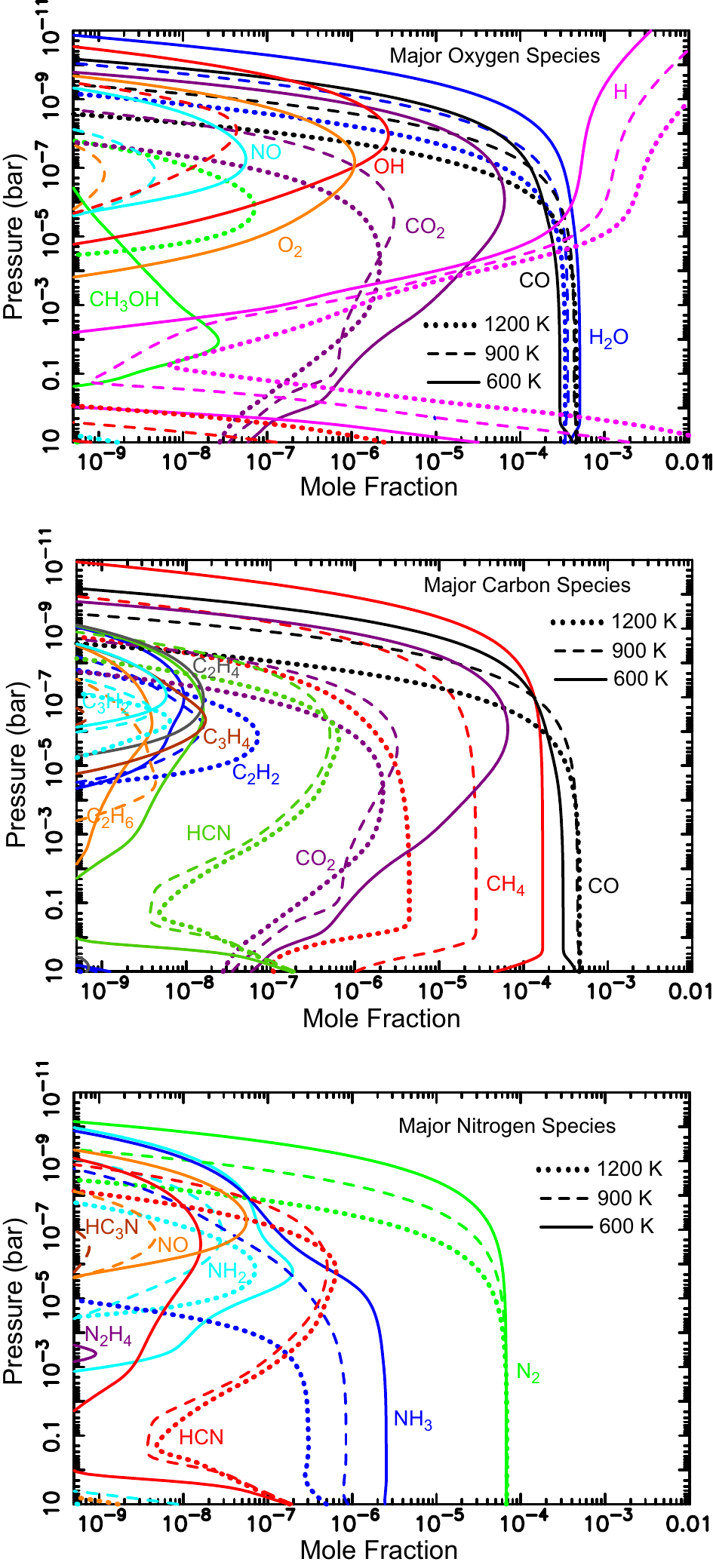}
\caption{The vertical mixing-ratio profiles of several atmospheric species as a function of \teff\ for 
a planet with $g$ = 10$^{3.5}$ cm s$^{-2}$ and \kdeep\ = 10$^6$ cm$^2$ s$^{-1}$, that is being 
irradiated by an HR 8799-like star at a distance of 68 AU (dashed lines), for \teff\ = 1200 K (dotted lines), 
900 K (dashed lines), and 600 K (solid lines).  Most disequilibrium photochemical products are synthesized 
more effectively in low-\teff\ atmospheres, but some photochemical products (most notably HCN and C$_2$H$_2$) 
become more abundant at higher \teff. 
A color version of this figure is available in the online journal.\label{figmixtemp}}
\end{figure}

\subsubsection{Sensitivity of disequilibrium chemistry to temperatures}

Finally, many photochemical products on directly imaged planets tend to be very sensitive to temperature --- both the 
effective temperature of the planet, \teff\ (which on young Jupiters is controlled by the internal heat flux rather than 
radiation from the host star), and the temperature in the planet's stratosphere (i.e., the radiative region above the 
convecting troposphere).  Note that because irradiation from the host star has less of an effect than internal heat 
flow on the upper-atmospheric temperatures of these distant, young, hot, directly imaged planets, our generic young-Jupiter 
models with larger \teff\ have larger stratospheric temperatures, too (see Fig.~\ref{figtemp}).  As discussed previously, 
\teff\ affects the quenched abundances of the photochemically active parent molecules, which can in turn influence the 
production rate of disequilibrium photochemical ``daughter'' products.  More importantly, the stratospheric temperatures 
affect the subsequent reaction rates of the photochemically produced molecules and radicals, as well as affect the height 
to which the photochemically active parent molecules are carried before molecular diffusion takes over and severely 
limits their abundance.  The altitude variation of this homopause level can change the pressure at which photolysis 
occurs, thereby affecting subsequent pressure-dependent reactions.  Figure \ref{figmixtemp} shows how the vertical 
profiles of some of the major photochemically active molecules in our models vary with temperature, for planets 
with \teff\ = 600, 900, or 1200 K, and log($g$) = 3.5 (cgs), \kdeep\ = 10$^6$ cm$^2$ s$^{-1}$, orbiting at 68 AU from 
a star with the properties of HR 8799.  Although variations 
in \teff\ have a relatively straightforward influence on the quenched species' abundances, the response to upper atmospheric 
temperatures is more complicated.

Smaller \teff\ results in larger quenched abundances of CH$_4$, NH$_3$, and H$_2$O (all other factors being equal), and 
allows these molecules to be carried to higher homopause altitudes, so one might naively assume that these factors lead 
to greater abundances of photochemical products on cooler planets.  However, photolysis in these young-Jupiter models 
is photon-limited rather than species-limited, and the column-integrated photolysis rate of water --- which produces H, 
as well as OH, and thus drives much of the subsequent photochemistry for carbon, nitrogen, and oxygen species --- is 
only slightly different for all three different \teff\ models shown in Fig.~\ref{figmixtemp}. Instead, the critical factor 
is the efficiency of recycling of the parent species versus competing reactions to form other products.  When temperatures 
are larger, recycling of water is more prevalent through reactions such as OH + H$_2$ $\rightarrow$ H$_2$O + H, which has 
a high energy barrier and operates more effectively at high temperatures.  Therefore, fewer reactive OH and O radicals are
available to form oxygen-rich photochemical products such as CO$_2$, H$_2$CO, CH$_3$OH, or O$_2$ when temperatures are 
higher \citep[see also][]{zahnle16}.  Moreover, the H atom abundance increases as the upper-atmospheric temperature 
increases (due to the more efficient catalytic destruction of H$_2$ following water photolysis), and the increased H atom 
abundance decreases the stability of some photochemical products such as CO$_2$ and C$_2$H$_6$.  

On the other hand, the more efficient atomic H production at 
high temperatures leads to an overall increase in the production rate of reactive CH$_3$ and NH$_2$ radicals as the temperature 
increases, as a result of reactions like H + $\CHfour$ $\rightarrow$ $\CHthree$ + H$_2$ and H + NH$_3$ $\rightarrow$ NH$_2$ + H$_2$, 
and even though the reverse recycling reactions are also more effective at high temperatures, the nitrogen- and carbon-bearing 
products can still form at any temperature.  The result is that some photochemical products, like HCN and C$_2$H$_2$ that 
have strong bonds and are more stable at high temperatures, are produced more efficiently at higher \teff, while 
other species like C$_2$H$_6$, C$_3$H$_4$, and N$_2$H$_4$ are produced more efficiently at lower \teff.  The peak production 
altitude and overall shape of the mixing-ratio profiles can vary with \teff, as well (see Fig.~\ref{figmixtemp}).

As emphasized by \citet{zahnle16}, the oxygen-bearing photochemical products are particularly sensitive to the 
upper-atmospheric temperature, and the abundance of the oxygen species increases significantly when stratospheric 
temperatures fall below $\sim$250 K.  The rate coefficient for the water recycling reaction OH + H$_2$ $\rightarrow$ 
H$_2$O + H drops by almost three orders of magnitude with a reduction in temperature from 500 K to 200 K \citep{baulch05}.  
The reduced efficiency of OH + H$_2$ $\rightarrow$ H$_2$O + H at low temperatures opens the door for efficient carbon 
oxidation, and CO + OH $\rightarrow$ CO$_2$ + H becomes a competitive loss process for the OH.  As a result, neither 
H$_2$O nor CO are as efficiently recycled in the colder atmospheres, and the OH + CO reaction will proceed 
effectively until it depletes enough CO that the OH + H$_2$ reaction can again compete as a loss process for the OH.  
One then sees a depletion of H$_2$O and CO at high altitudes in the coldest models, with a concomitant increase in 
CO$_2$ and other oxygen products like O$_2$ and CH$_3$OH that can form when OH does not effectively recycle back 
to water.  Carbon dioxide becomes a spectroscopically significant photochemical product on colder young Jupiters 
(see section~\ref{sectgenspec}), and the effect is further magnified the greater the incident UV flux.

\begin{figure*}
\begin{center}
\includegraphics[scale=0.8]{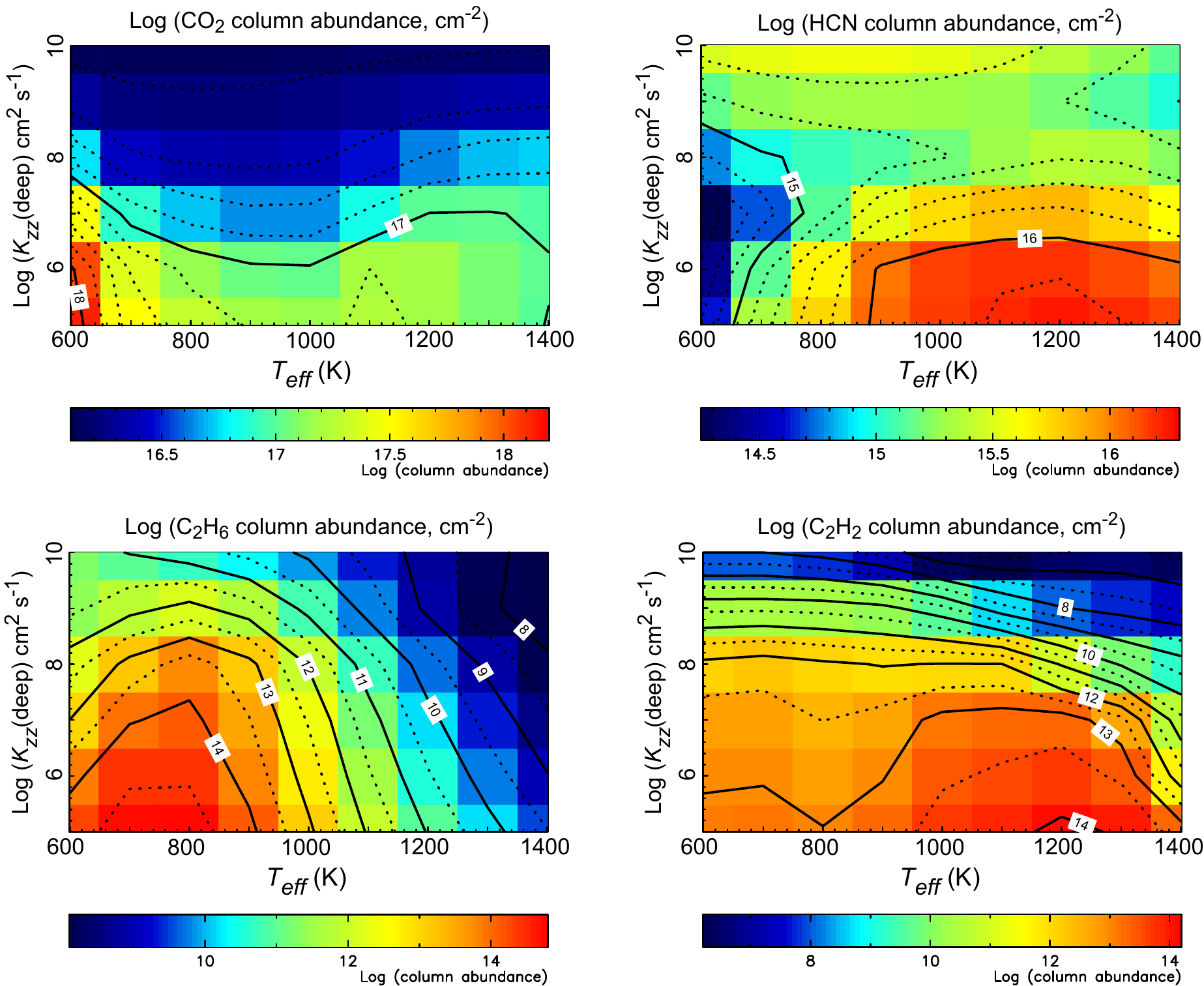}
\caption{Integrated column abundance of CO$_2$ (top left), HCN (top right), C$_2$H$_6$ (bottom left), and C$_2$H$_2$ 
(bottom right) above 1 mbar as a function of \teff\ and \kdeep\ for planets with a surface gravity of $g$ = 10$^{3.5}$ 
located at 68 AU from a star with the properties of HR 8799.  Photochemistry dominates in this region of the atmosphere, 
and different species exhibit a complicated sensitivity to both \teff\ and \kdeep.
A color version of this figure is available in the online journal.\label{figchemgrid}}
\end{center}
\end{figure*}

Figure \ref{figchemgrid} provides further details showing how the photochemical products CO$_2$ (top left), 
HCN (top right), C$_2$H$_6$ (bottom left), and C$_2$H$_2$ (bottom right) vary with 
changes in both \teff\ and \kdeep, for  planets with log($g$) = 3.5 cgs located 68 AU from a star like HR 8799.  
For the shape of the 
vertical \kzz\ profiles we have assumed (see Fig.~\ref{figeddy}), smaller \kdeep\ values also correspond to weaker 
eddy mixing in the lower stratosphere, which increases the residence time for photochemical products synthesized at 
higher altitudes, allowing them to build up to larger abundances.  Therefore, most photochemical products exhibit 
increased abundances for smaller \kdeep\ values.  One exception is HCN, which has a more complicated dependence 
because larger \kdeep\ values favor larger quenched abundances of HCN; i.e., quenching, not just photochemistry, 
contributes to the overall abundance of HCN.  For any particular \kdeep\ value, the temperature dependence can be 
complicated, with CO$_2$ exhibiting a major increase at the lowest temperatures for the reasons discussed above, 
C$_2$H$_6$ being favored at moderately low temperatures, and C$_2$H$_2$ and HCN being favored at \teff\ $\approx$ 1200 K.

In general, hydrocarbons such as C$_2$H$_6$ and C$_2$H$_2$ are not expected to become abundant enough to be observable 
on young Jupiters, except potentially for closer-in planets (i.e., those receiving a large UV flux) in combination with 
a more stagnant (lower \kzz) lower stratosphere and an increasingly well-mixed and colder ($\lta$250 K) upper stratosphere, 
in which water recycling is less effective and the resulting H production is reduced.  Low upper-atmospheric 
temperatures favor C$_2$H$_6$ over C$_2$H$_2$, while higher temperatures favor C$_2$H$_2$.  The quenched HCN abundance 
reaches potentially observable abundances of a few $\times$ 10$^{17}$ cm$^{-2}$ above 100 mbar for large \kdeep\ 
($\gta$ 10$^{9}$ cm$^{2}$ s$^{-1}$), and a high UV flux combined with moderate \teff\ of 1100--1300 K would provide 
an increased photochemical component on top of that that quenched HCN.  Carbon dioxide is the big winner from a 
disequilibrium-chemistry standpoint, with observable quantities (see section~\ref{sectgenspec}) of greater than 10$^{18}$ 
cm$^{-2}$ above 100 mbar being produced through both quenching and photochemistry in all the models studied, with a 
column abundance greater than 10$^{19}$ cm$^{-2}$ above 100 mbar forming in the planets with cooler, more stagnant 
lower stratospheres.  

\begin{figure*}
\begin{tabular}{ll}
{\includegraphics[angle=0,clip=t,scale=0.42]{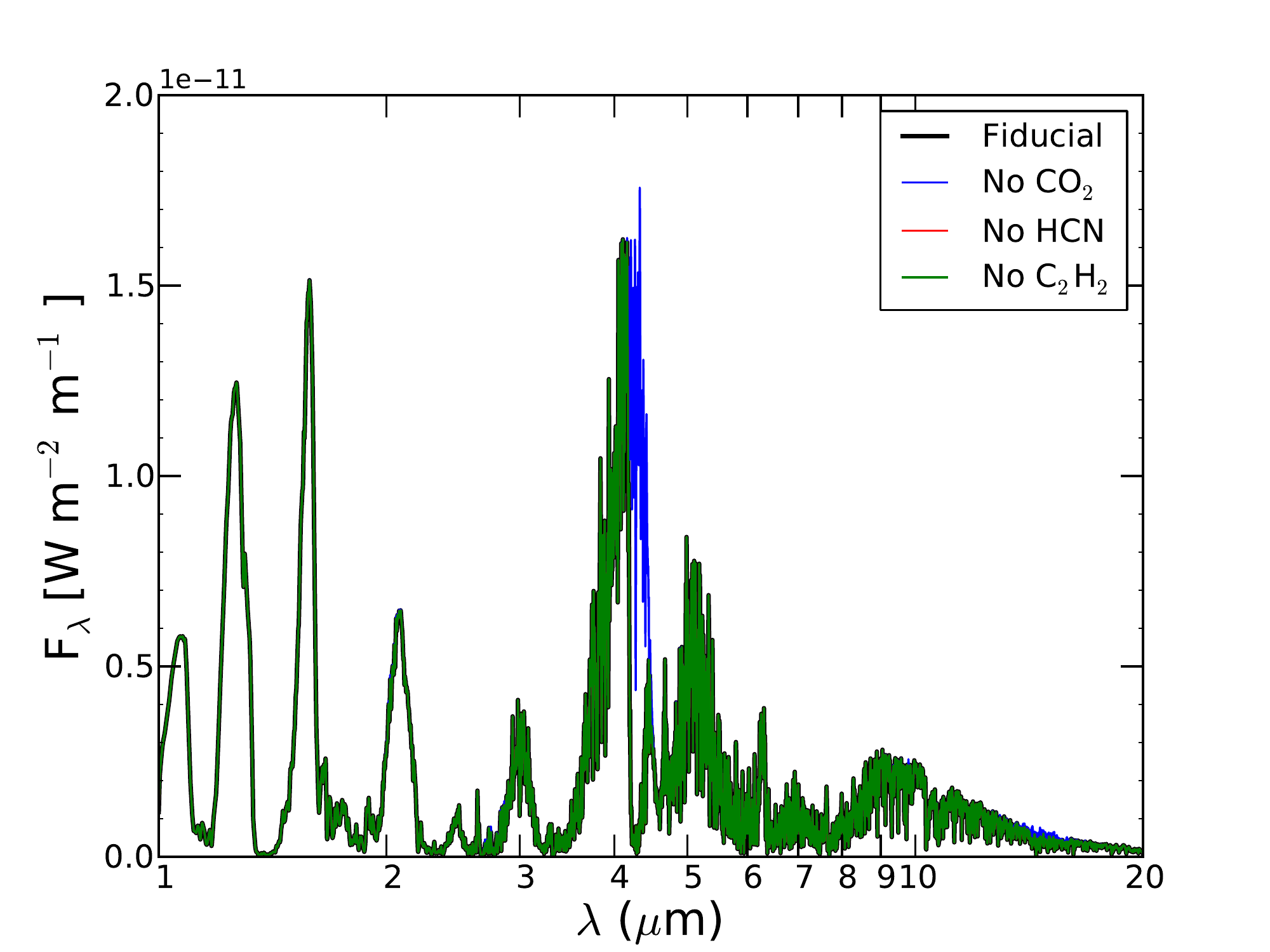}}
&
{\includegraphics[angle=0,clip=t,scale=0.42]{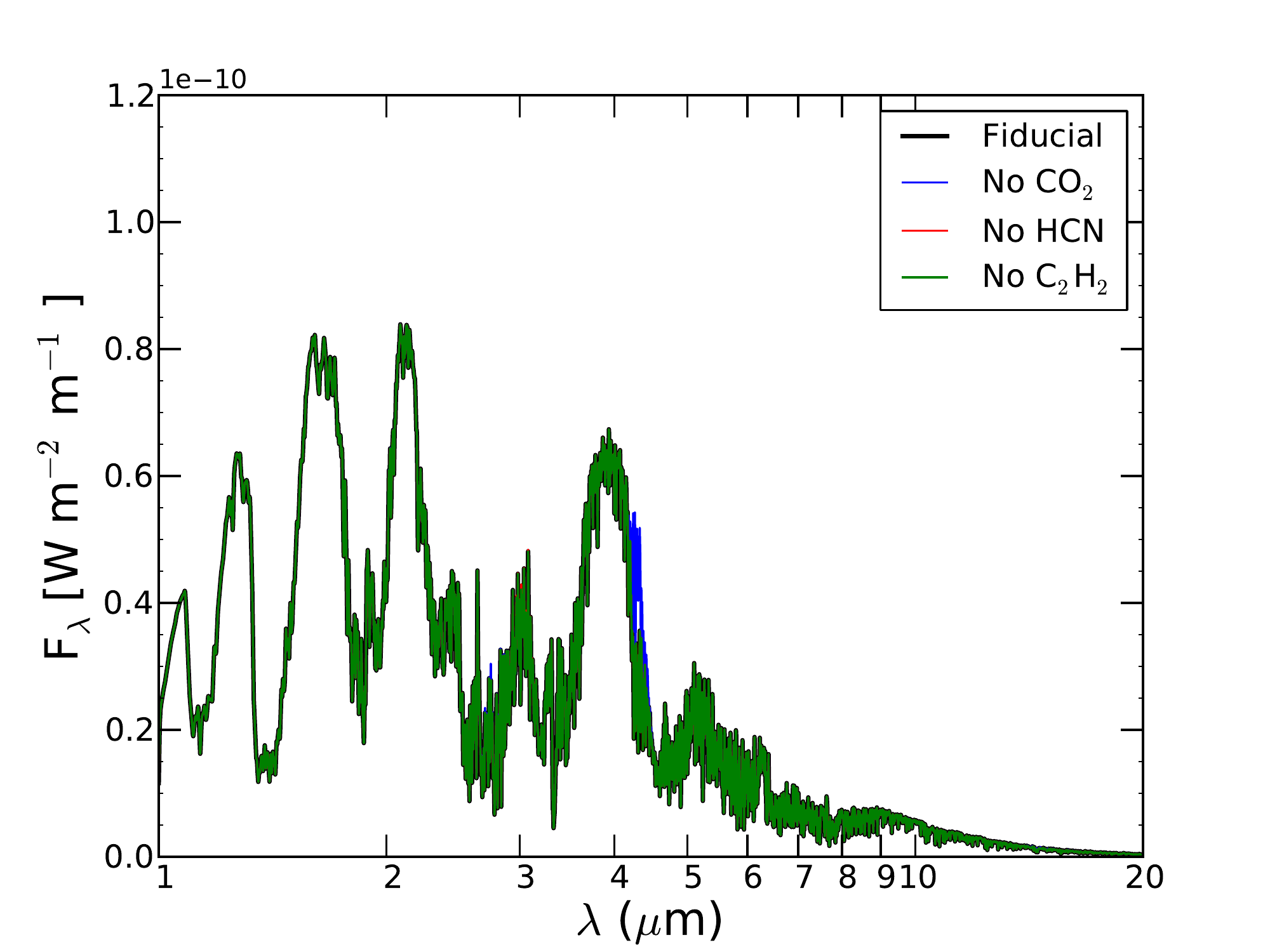}}
\\
\end{tabular}
\caption{Synthetic spectra from our photochemical models of generic young Jupiters orbiting at 68 AU from 
a star with the properties of HR 8799, with planetary properties of log($g$) = 4.0 cgs, 
$R$ = 1.2$R_J$, a global gray absorbing cloud (no patchiness), at a distance of 39 pc from Earth, for (Left) \teff\ = 
600 K and \kdeep\ = 10$^5$ cm$^2$ s$^{-1}$ and (Right) \teff\ = 1000 K and \kdeep\ = 10$^7$ cm$^2$ s$^{-1}$.
The cloud base is assumed to be located at the pressure where the MgSiO$_3$ condensation curve crosses the 
temperature profile, and the cloud is assumed to extend to the top of the atmosphere, with the opacity 
adjusted such that the optical depth is unity between 1 and 10$^{-4}$ bars.  The plots show how various 
photochemical products affect the spectra, through the removal of CO$_2$ (blue), HCN (red), and C$_2$H$_2$ 
(green) from the spectral calculations.  Of these photochemical products, only CO$_2$ affects the spectra 
significantly at near-IR wavelengths.
A color version of this figure is presented in the online journal.\label{specgenericco2}}
\end{figure*}

\begin{figure}
\includegraphics[scale=0.44]{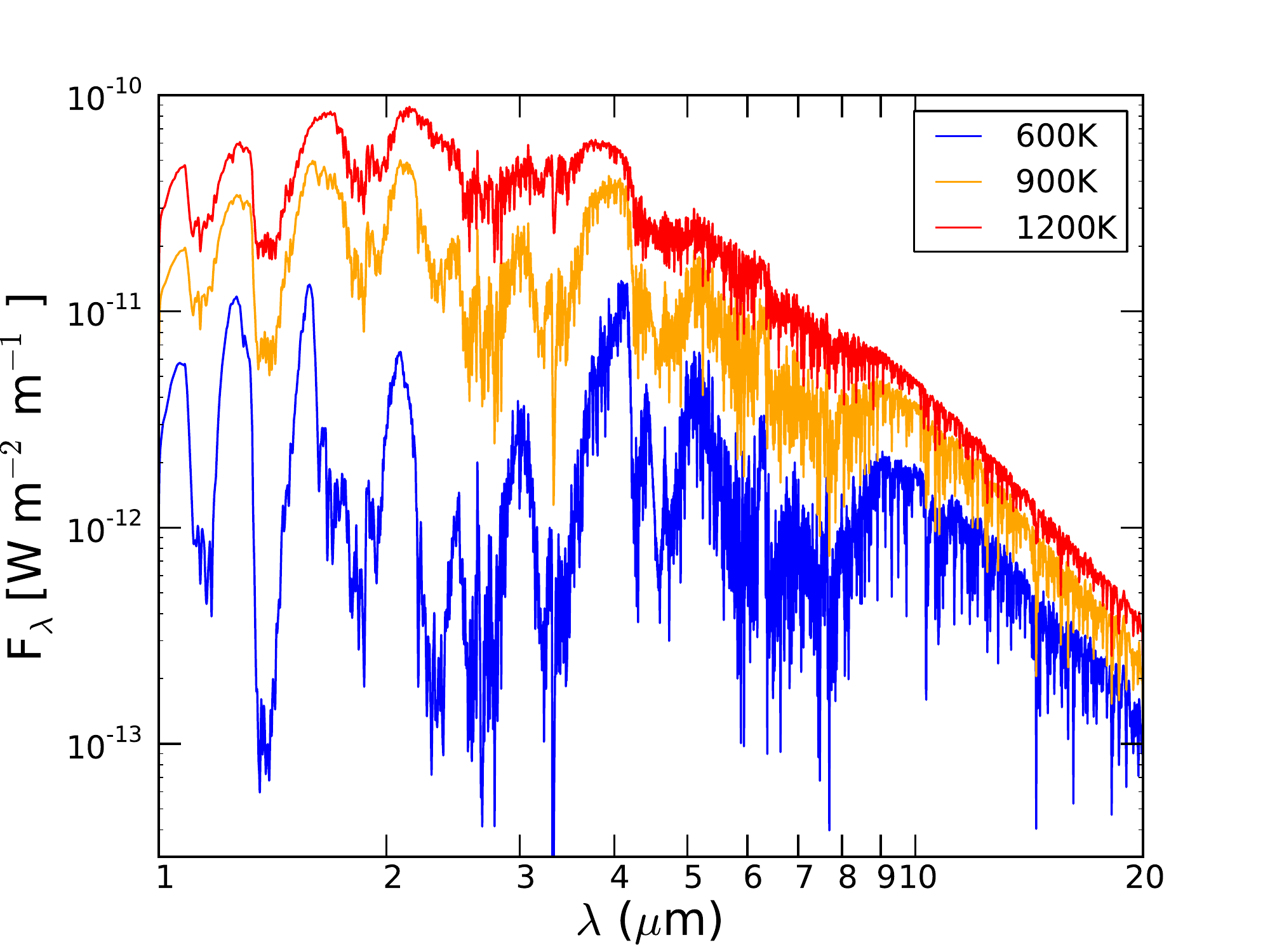}
\caption{Synthetic spectra from our photochemical models of generic young Jupiters orbiting 68 AU from a star 
with the properties of HR 8799, with surface gravities $g$ = 3200 cm s$^{-2}$, eddy \kdeep\ = 10$^6$ cm$^2$ s$^{-1}$, 
and effective temperatures \teff\ = 600 K (blue), 900 K (orange), and 1200 K (red).  These models correspond to 
the ones shown in Fig.~\ref{figmixtemp}.  For the purpose of the spectral calculations, we have assumed that 
the planets have radii = 1.0$R_J$, are located 39 pc from Earth, and possess uniform gray absorbing clouds with 
optical depths of one between the base of the MgSiO$_3$ condensation region and the top of the atmosphere.  
Note that the absorption in most of the molecular bands (e.g., H$_2$O, CH$_4$, NH$_3$, and CO$_2$) increases as \teff\ 
decreases (cf.~Fig.~\ref{figmixtemp}).
A color version of this figure is available in the online journal.\label{specvsteff}}
\end{figure}

\begin{figure}
\includegraphics[scale=0.44]{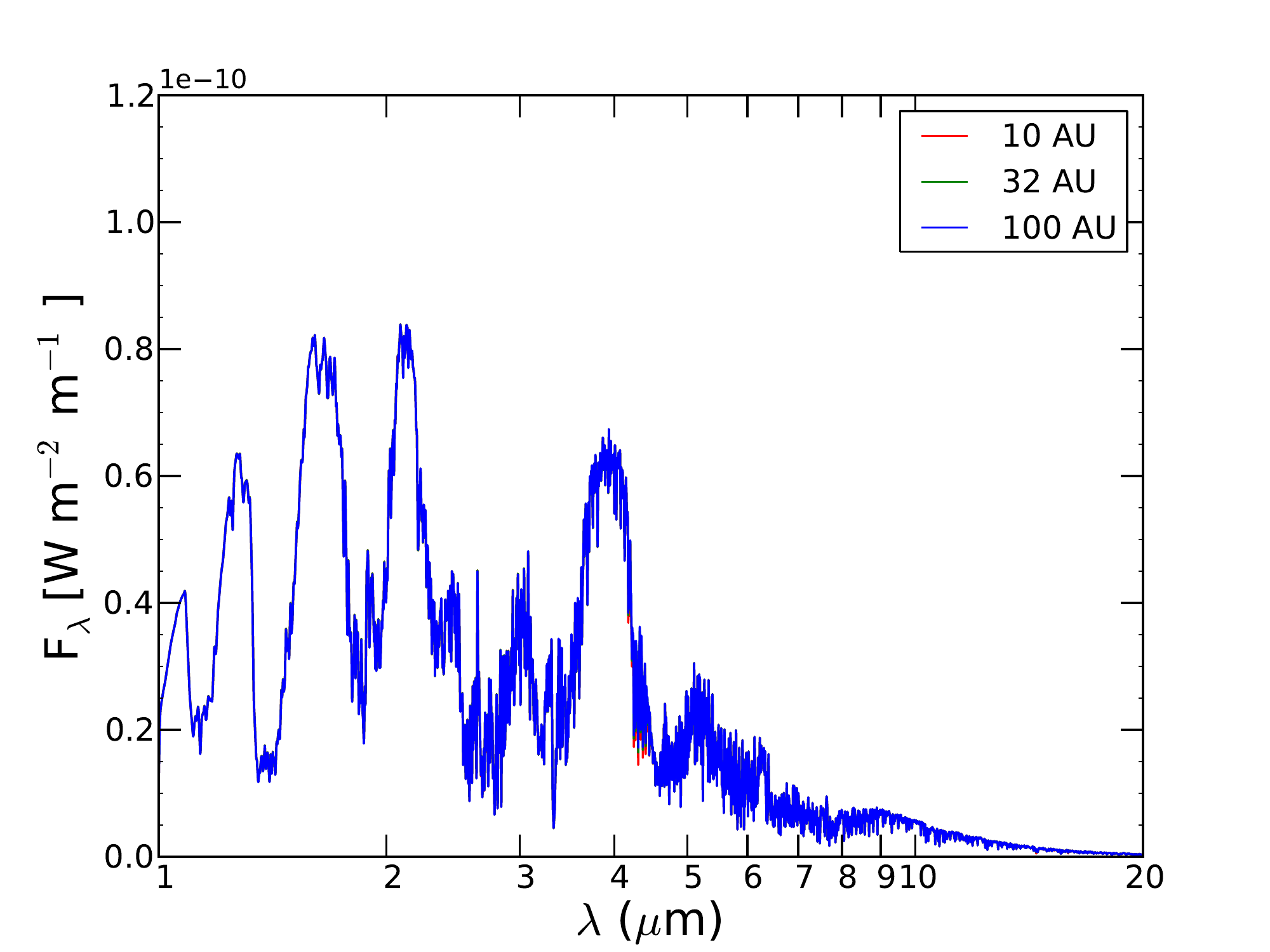}
\caption{Synthetic spectra from our photochemical models of generic young Jupiters with \teff\ = 1000 K, $g$ = 
10$^4$ cm s$^{-2}$, \kdeep\ = 10$^7$ cm$^2$ s$^{-1}$, orbiting a star with the properties of HR 8799 at 
10 AU (red), 32 AU (green), and 100 AU (blue).  These models correspond to the ones shown in Fig.~\ref{figdistance}.  
For the purpose of the spectral calculations, we have assumed that the planets have radii = 1.2$R_J$, are located 
39 pc from Earth, and possess uniform gray absorbing clouds with optical depths of one between 1 and 10$^{-4}$ mbar.  
Note that at this relatively high \teff\ and \kdeep\ the photochemical products have little impact on the 
spectrum, except for the minor increase in CO$_2$ absorption at 4.2-4.3 $\mu$m in the shorter-period model, 
due to its greater photochemical production and correspondingly larger CO$_2$ abundance.
A color version of this figure is available in the online journal.\label{specvsdistance}}
\end{figure}

\subsection{Generic Directly Imaged Planets: Spectra\label{sectgenspec}}

We show selected spectra from our directly imaged planets in Figs.~\ref{specgenericco2} \& 
\ref{specvsteff}.  These synthetic spectra were generated from the forward radiative-transfer model 
described in \citet{line13ret,line14dwarf,line15dwarf}.  First, Fig.~\ref{specgenericco2} shows results from 
two generic models with different quenched abundances of CH$_4$ and CO.  Both planets are assumed to be 39 pc 
from Earth, with surface gravities of 10$^4$ cm s$^{-2}$, a radius of 1.2$R_J$, and a uniform gray absorbing aerosol 
layer with a base located where the thermal profile crosses the MgSiO$_3$ condensation curve and a total optical depth 
of unity between 1 bar and 10$^{-4}$ bars.  Both planets are assumed to orbit 68 AU from a star with properties of 
HR 8799.  The planet shown in the left panel has \teff\ = 600 K and \kdeep\ = 10$^5$ 
cm$^2$ s$^{-1}$, for which the quenched CH$_4$ mixing ratio is 2.4 times that of CO (see Fig.~\ref{figgridall}).  
The planet in the right panel has \teff\ = 1000 K and \kdeep\ = 10$^7$ cm$^2$ s$^{-1}$, such that the quenched CO 
mixing ratio is 18 times that of CH$_4$.  Absorption features of H$_2$O are readily apparent in the spectra 
of both planets in bands near $\sim$1.4, $\sim$1.8-1.9, $\sim$2.6-2.8, and the $\sim$5.5-7.5 $\mu$m region, and 
CO absorption features are apparent in both plots in the $\sim$4.5-4.8 $\mu$m region.  Although CH$_4$ absorption 
features are also obvious in both plots, the bands at 2.3, 3.3, and 7.7 $\mu$m are deeper for the cooler planet, with  
its larger quenched methane abundance.  The cooler planet also has a larger column of photochemically produced 
CO$_2$, which shows up most distinctly in the 4.2-4.3 $\mu$m absorption bands on both planets, as well as more 
subtlely in the 2.7-2.8-$\mu$m region on the warmer planet and the $\sim$14-16 $\mu$m region on the cooler planet.  
Absorption in the 4.2--4.3-$\mu$m CO$_2$ bands should be particularly apparent on young Jupiters, trending toward 
greater absorption for lower \teff.  HCN is abundant enough on the warmer, more rapidly mixed planet 
(see Fig.~\ref{figchemgrid}) to have a minor effect on the spectrum at 3 $\mu$m, while other photochemical 
products such as C$_2$H$_2$ are not abundant enough to notably affect the spectra for either of these generic 
young Jupiters considered.

Figure \ref{specvsteff} further illustrates how the spectra of our generic young Jupiters changes as 
a function of \teff.  In this figure, we plot the synthetic spectra from the photochemical models shown in 
Fig.~\ref{figmixtemp} --- these planets are assumed to orbit 68 AU from a star with properties similar to HR 8799, and 
have $g$ = 3200 cm s$^{-2}$, \kdeep\ = 10$^6$ cm$^2$ s$^{-1}$, and \teff\ = 600, 900, or 1200 K.  For the 
spectral calculations, we again assume that the systems are located 39 pc from Earth, with planetary 
radii = 1.0$R_J$ and uniform gray absorbing clouds with optical depths of one between the base of 
the MgSiO$_3$ condensation region and the top of the atmosphere.  The cooler planet contains more quenched H$_2$O 
and CH$_4$ and possesses a colder stratosphere, so the absorption bands due to these species are therefore deeper.  
The cooler planet also has more quenched NH$_3$, which shows up readily near 10.4 $\mu$m, and more photochemically 
produced CO$_2$, which is notable in the 4.2-4.3 and $\sim$15 $\mu$m bands. 

The influence of photochemically produced CO$_2$ on the emission spectrum of young Jupiters diminishes 
strongly with increasing \teff\ and increasing \kzz\ in the stratosphere, and is, in particular, highly 
sensitive to the stratospheric temperature, as discussed in section 3.1.4.  For sufficiently large stratospheric 
\kzz\ and temperatures, an increased UV irradiation 
level does not overcome the tendency toward small overall CO$_2$ column abundances.  For example, 
Fig.~\ref{specvsdistance} shows how spectra from the \teff\ = 1000 K, log($g$) = 4.0, \kdeep\ = 10$^7$ cm$^2$ s$^{-1}$ 
models from Fig.~\ref{figdistance} vary with orbital distance ranging from 10, 32, and 100 AU.  The spectra are 
similar for all three planets.  There is a slight difference in the 4.2-4.3 $\mu$m region due to increased CO$_2$ 
absorption for the shorter-period planets, but these differences are small.  
In general, the spectra of young directly imaged giant planets will be dominated by quenched H$_2$O, 
CH$_4$, and CO, but absoption features due to photochemically produced species such as CO$_2$ can be important 
when \teff\ is small, lower-stratospheric eddy mixing coefficients are small (which allow larger column 
abundances of photochemical species to build up), and UV irradiation levels are large.

\begin{deluxetable*}{llll}
\tabletypesize{\scriptsize}
\tablecaption{Column abundances for the HR 8799~\lowercase{b} models\label{tabcolumnb}}
\tablewidth{400pt}
\tablecolumns{4}
\tablehead{
\colhead{ } & \colhead{Column abundance} & \colhead{Column abundance} & \colhead{Column abundance} \\
\colhead{Species} & \colhead{above 10 mbar} & \colhead{above 100 mbar} & \colhead{above 1 bar} \\
\colhead{ } & \colhead{(cm$^{-2}$)} & \colhead{(cm$^{-2}$)} & \colhead{(cm$^{-2}$)} 
}
\startdata
CH$_4$       & (2--3)$\scinot18.$      & (2--3)$\scinot19.$      & (2--3)$\scinot20.$      \\
C$_2$H$_2$   & (0.01--5)$\scinot12.$   & (0.01--5)$\scinot12.$   & (0.8--8)$\scinot14.$    \\
H$_2$O       & (1--3)$\scinot20.$      & (1--3)$\scinot21.$      & (1--3)$\scinot22.$      \\
CO           & (2.8--4.6)$\scinot20.$  & (2.8--4.5)$\scinot21.$  & (2.8--4.5)$\scinot22.$  \\
CO$_2$       & (0.5--4)$\scinot17.$    & (0.5--3)$\scinot18.$    & (0.4--1)$\scinot19.$    \\
NH$_3$       & (2.5--3)$\scinot17.$    & (2.5--3)$\scinot18.$    & (2.5--3)$\scinot19.$    \\
HCN          & (1--3)$\scinot16.$      & (1--2.5)$\scinot17.$    & (1--2.5)$\scinot18.$    \\
\enddata
\tablecomments{Models assume log($g$) = 3.47--3.5 cgs.}
\end{deluxetable*}

\subsection{HR 8799 b\label{sect8799b}}

Of the four planets detected in the HR 8799 system \citep{marois08,marois10}, HR 8799 b is the farthest away from the host 
star \citep[68 AU,][]{maire15} and seems to be the smallest and coolest \citep[e.g.,][]{marois08}.  Most comparisons 
of spectral models with observational data favor \teff\ in the broad range 700--1200 K and log($g$) = 3.0-4.5 cgs for HR 8799 b 
\citep{marois08,hinz10,barman11hr8799,barman15,currie11,currie14,galicher11,madhu11hr8799,marley12,skemer12,ingraham14,rajan15}.  
The broad range here stems from degeneracies between \teff, log($g$), assumed cloud properties, planetary radius, and metallicity.  
Moreover, the models tend to have difficulty simultaneously fitting the short-wavelength infrared spectra (1--2.5 $\mu$m), which 
show evidence for deeper molecular absorptions, and the longer-wavelength  mid-infrared photometric (3--5 $\mu$m), which exhibit 
flatter spectral behavior.  These difficulties complicate the derivation of planetary properties.  The best-fit models typically 
seem to require thick but patchy clouds, and the spectrum of HR 8799 b is distinctly different from brown dwarfs with the same 
effective temperature.

For our HR 8799 b  models, we adopt the recent constraints of \citet{barman15} (\teff\ = $1000 \pm 100$ K 
and log($g$) = 3.5 $\pm$ 0.5 cgs) because their analysis of the medium-resolution $H$- and $K$-band data with the OSIRIS 
instrument at Keck 
have provided the best available constraints on the abundances of CH$_4$, H$_2$O, and CO.  For consistency with the \citet{barman15}
modeling procedure and their preferred restriction of C and O abundances to possible sequences derived from the \citet{oberg11} disk 
chemical evolution model, we also adopt a slightly super-solar C/O ratio of 0.65--0.7 for these models, and metallicities of 
$\sim$0.6-1.0 times solar.  

\begin{figure}[htb]
\includegraphics[scale=0.7]{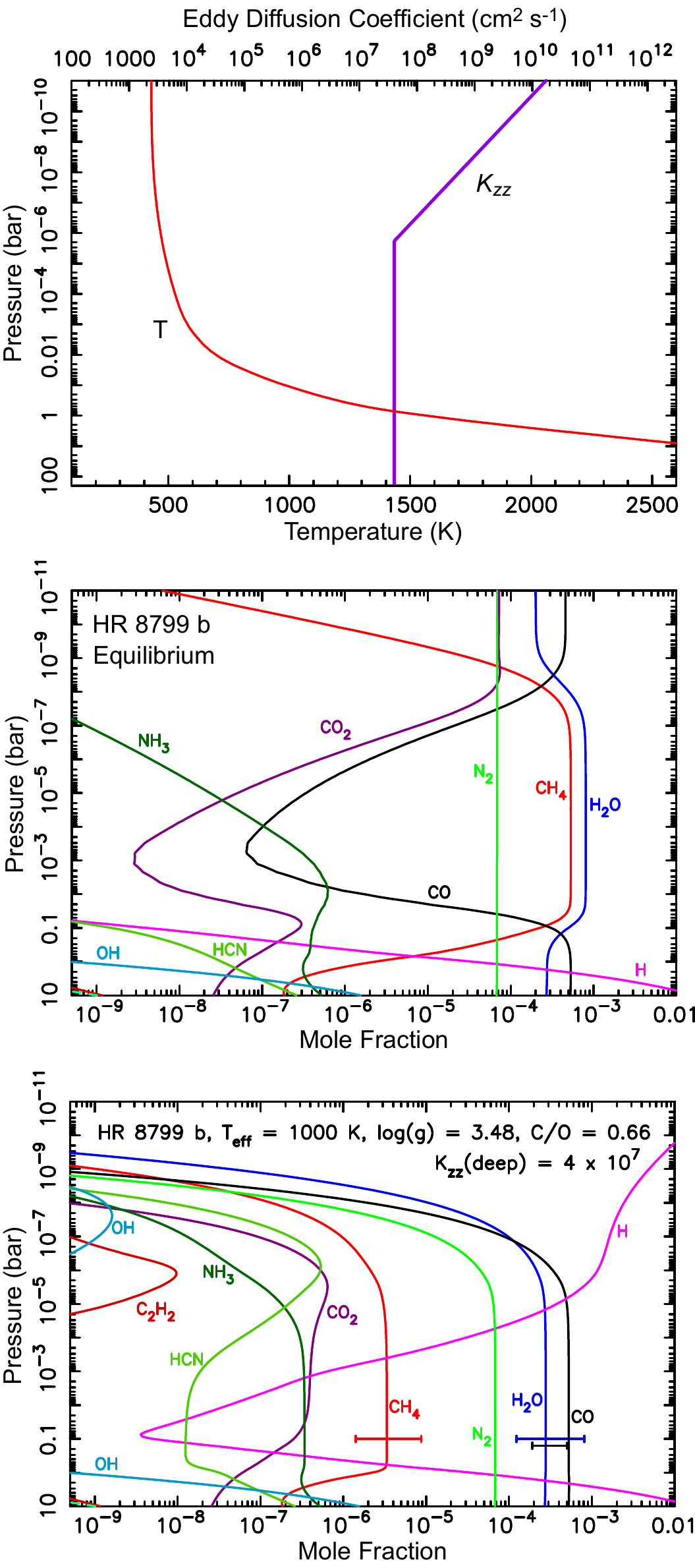}
\caption{Chemical model for HR 8799 b assuming \teff\ = 1000 K, $g$ = 3000 cm s$^{-2}$ (assumed $M$ = 1.9$M_J$), and solar 
metallicity, except C/O = 0.66: 
(Top) The temperature profile (red curve, bottom axis) from the radiative-convective equilibrium model of \citet{marley12} assuming 
the above bulk constraints, and the eddy diffusion coefficient profile (purple curve, top axis) adopted in the photochemical model; 
(Middle) the predicted thermochemical equilibrium mixing-ratio profiles for the major oxygen, carbon, and nitrogen species, as labeled, 
for the assumed pressure-temperature profile; (Bottom) mixing-ratio profiles predicted from our thermo/photochemical kinetics and 
transport model for the above thermal structure, \kzz\ profile, and assumed bulk elemental composition.  The line segments in the 
bottom plot are the observational constraints for CH$_4$ (red), H$_2$O (blue), and CO (black) from \citet{barman15}.
A color version of this figure is available in the online journal.\label{hr8799b_marley}}
\end{figure}

Fig.~\ref{hr8799b_marley} shows the results from one of our HR 8799 b models.  In this model, we have assumed \teff\ = 1000 K, 
$g$ = 3000 cm s$^{-2}$ (with assumed mass 1.9$M_J$), and a solar metallicity atmosphere except for a C/O ratio of 0.66, and 
we have used the radiative-convective equilibrium model of \citet{marley12} to define the temperature structure.  With this 
thermal structure, the quenched CH$_4$ abundance falls within the 1.4$\scinot-6.$--8.7$\scinot-6.$ mixing-ratio
constraints provided by \citet{barman15} when log(\kdeep) 
$\approx$ 6--9, with a best fit for \kdeep\ = $4\, \times \, 10^{7}$ cm$^2$ s$^{-1}$.  Figure~\ref{hr8799b_marley} 
demonstrates that the CO mixing ratio is expected to be much larger than the CH$_4$ mixing ratio on HR 8799 b as a result of 
transport-induced quenching.  Similarly, the quenched N$_2$ abundance is much greater than that of NH$_3$, and H$_2$O 
quenches at a mixing ratio a factor of $\sim$3 smaller than equilibrium predictions.  As expected (see 
section \ref{sectgeneric}), the CO$_2$ and HCN abundances are also significantly enhanced in comparison to chemical equilibrium as 
a result of quenching of the dominant oxygen, carbon, and nitrogen species \citep[see also][]{moses11}.  The coupled carbon-oxygen 
and carbon-nitrogen photochemistry described in section \ref{sectgeneric}) leads to an additional peak in the CO$_2$ and HCN 
abundances at high altitudes, which for the case of HCN adds notably to the stratospheric column abundance.  Hydrocarbons such as 
C$_2$H$_2$ and C$_2$H$_6$ and key radicals such as OH and NH$_2$ are produced from high-altitude photochemistry, but these species
are less stable in the lower stratosphere, and they never reach observable column abundances.  

Overall, although disequilibrium 
quenching is very important in controlling the atmospheric composition of HR 8799 b --- including controlling the abundance of minor 
species not typically considered in simple quenching models --- photochemistry itself is less important due to relatively warm 
stratospheric temperatures (which tend to decrease the stability of photochemical products) and the mild UV flux received by 
HR 8799 b.  If the lower-stratospheric eddy $K_{zz}$ values were much lower than we have assumed here, then the column abundance 
of key photochemical products like C$_2$H$_x$ hydrocarbons could be increased, although it is still unlikely that they could 
achieve observable values.

\begin{figure}[htb]
\includegraphics[scale=0.5]{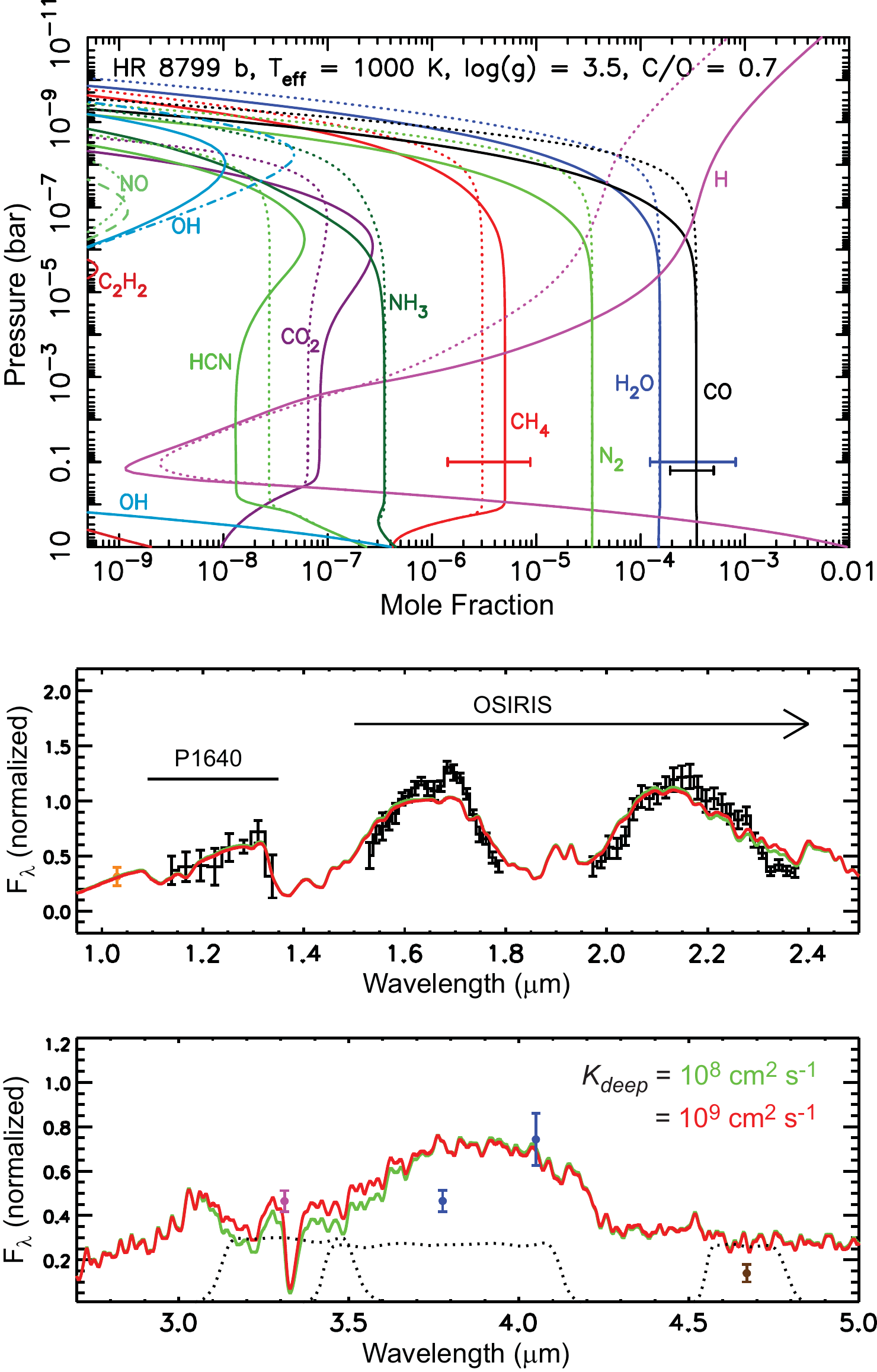}
\caption{(Top) Model results for HR 8799 b assuming \teff\ = 1000 K, log($g$) = 3.5 cgs, C/O = 0.7, 
a subsolar metallicity, a temperature profile from \citet{barman15}, and \kdeep\ = 10$^8$ cm$^2$ $\smone$ 
(solid curves) and 10$^9$ cm$^2$ $\smone$ (dotted curves). (Bottom two panels) HR 8799 b observations (black data 
points with error bars; see text) compared with synthetic spectra generated from our thermo/photochemical kinetics and
transport models from the top panel, for \kdeep\ = 10$^8$ cm$^2$ $\smone$ (green) and 10$^9$ cm$^2$ $\smone$ (red); 
see text and \citet{barman15} for details.
A color version of this figure is available in the online journal.\label{hr8799b_barman}}
\end{figure}

Other HR 8799 b models developed with different assumptions about the thermal structure and other planetary parameters 
produce similar results.  For colder (warmer) thermal structures, it takes a larger (smaller) \kdeep\ to quench CH$_4$ 
at the same abundance as in the above model.  As an example, Fig.~\ref{hr8799b_barman} shows the results for two photochemical 
models that assume \teff\ = 1000 K, $g$ = 3162 cm s$^{-2}$, a C/O ratio of 0.7, a subsolar metallicity (i.e., $\sim$0.63 
times the solar O/H of \citealt{grevesse07}), a thermal structure that is taken from \citet{barman15}, and an assumed \kdeep\ 
that is 10$^8$ cm$^2$ $\smone$ (solid curves) or 10$^9$ cm$^2$ $\smone$ (dotted curves).  These models are cooler everywhere 
than the one shown in Fig.~\ref{hr8799b_marley}, and so it takes a larger \kdeep\ to quench CH$_4$ at the same abundance as 
the previous model.  If the eddy diffusion coefficient were to remain high in the stratosphere, as in the models shown here, 
then the photochemical species produced at high altitudes could diffuse rapidly through the stratosphere to deeper, warmer 
levels, where they would readily be converted back to the major quenched species. So again, photochemistry does not have 
much of an effect on the spectroscopically active molecules for these HR 8799 b models.  However, transport-induced quenching 
does play a major role in shaping atmospheric composition, including for the species H$_2$O, CO, CH$_4$, CO$_2$, N$_2$, NH$_3$, 
and HCN.  Quenching on a lower-gravity planet readily explains why the observed 
CH$_4$ absorption is so much less significant on HR 8799 b than on brown dwarfs of similar effective temperatures 
\citep[see also][]{zahnle14,barman11hr8799,barman15}.  Although \kdeep\ can in theory be constrained by comparing disequilibrium 
models like these to observations, in practice the thermal structure of the planet is uncertain enough that firm 
constraints are not possible.  We simply conclude that the deep-atmospheric mixing is strong (\kdeep\ $>$ 10$^7$ cm$^2$ s$^{-1}$) 
on HR 8799 b, consistent with that expected from convection on a planet with a strong internal heat source \citep[e.g.,][]{stone76}. 

Fig.~\ref{hr8799b_barman} also shows how synthetic spectra from these models compare to actual observations. 
In order of shorter to longer wavelengths in this figure, we plot the z/Y-band flux of \citet{currie11} 
in orange, the low-resolution P1640 \textit{J}-band spectrum of \citet{oppenheimer13} in black, 
the \textit{H}-band OSIRIS spectrum of \citet{barman11hr8799} in black, the \textit{K}-band OSIRIS spectrum 
of \citet{barman15} in black, and the longer-wavelength photometric data of \citet{skemer12} in pink, 
\citet{currie14} in blue, and \citet{galicher11} in brown. A 
small scaling was applied to the \textit{H}-band portion of the P1640 spectrum to get it to match the \textit{H}-band OSIRIS 
spectrum, as described in \citet{barman15}.  For other modeling assumptions, see \citet{barman15}. The quenched CH$_4$ abundance 
is sensitive to the assumed deep eddy diffusion coefficient \kdeep, and Fig.~\ref{hr8799b_barman} shows that in the near 
infrared, spectral observations in the 3.1-3.5 $\mu$m region are best suited to constraining the methane mixing ratio 
and hence \kdeep.  Although current ground-based and \textit{Hubble Space Telescope} observations can provide 
sufficient spectral information to loosely constrain CH$_4$ and thus \kdeep\ \citep[e.g.][]{konopacky13,barman15}, 
model degeneracies will be more easily broken with the broader wavelength coverage and moderate spectral resolution 
provided by the \textit{James Webb Space Telescope} (JWST) or other space-based instrumentation with coronographic 
spectroscopy capabilities.

The column abundances of key species in our full range of HR 8799 b models are listed in Table~\ref{tabcolumnb}.
Water is the dominant infrared opacity source and is readily detected in HR 8799 b spectra.  Methane and carbon monoxide 
have also been detected \citep[e.g.,][]{currie11,barman11hr8799,barman15}.
Tentative detections of NH$_3$ and/or $\CtwoHtwo$, and CO$_2$ or HCN have been reported by \citet{oppenheimer13} in 1.0--1.8 
$\mu$m spectra of the planet.  Many of these tentative detections are inconsistent with our HR 8799 b models.  For example, 
$\CtwoHtwo$ in our photochemical models never becomes abundant enough to be detectable on HR 8799 b for any of the 
infrared bands, including the relatively strong ones near 13.6 and $\sim$3 $\mu$m.  Carbon dioxide in the model is not 
abundant enough to be detectable in the 1--1.8 $\mu$m range, where the bands are weak, but it should be detectable in the stronger 
bands between 4--4.5 $\mu$m and near 15 $\mu$m; CO$_2$ may also be detectable in the $\sim$2.7--2.8 $\mu$m range if the photosphere 
extends down to $\sim$1 bar, but that may be problematic given that clouds are inferred to be present.  Hydrogen cyanide is 
potentially detectable in bands near 2.5, $\sim$3, and 6.8--7.4 $\mu$m if the photosphere extends deep, with a more likely 
stratospheric detection in the 14-$\mu$m band; however, HCN is not predicted to be abundant enough to be detectable in the 
1--1.8 $\mu$m region observed by \citet{oppenheimer13}.  Similarly, if the photosphere extends 
below $\sim$1 bar, NH$_3$ may be detectable near $\sim$1.5 $\mu$m, $\sim$2 $\mu$m, $\sim$3 $\mu$m, and $\sim$6.15 $\mu$m, 
but has the best chance of being detected in the stratosphere in the stronger bands in the 9--11 $\mu$m region.  Methane 
should be detectable in the $\sim$1.6 and 2.3 $\mu$m bands if the obscuring clouds are confined to altitudes below $\sim$100 mbar 
(and in fact CH$_4$ has been detected in the 2.3 $\mu$m band, \citealt{barman11hr8799,barman15}), with an even better 
chance of being detected in the stronger 3.3 $\mu$m band \citep[see][and Fig.~\ref{hr8799b_barman}]{currie11} and the 7.7 $\mu$m band.  The CO band in 
the 4.5--4.9 $\mu$m region should produce significant absorption in HR 8799 b spectra, and the band near 2.3--2.4 $\mu$m 
should also be observable \citep[see][]{barman15} and may help constrain cloud heights/thicknesses; however, moderate-resolution 
spectra are required, as some of the lines in this band overlap with H$_2$O and CH$_4$ lines, complicating identification 
\citep{barman15}.

\begin{figure}[htb]
\includegraphics[scale=0.7]{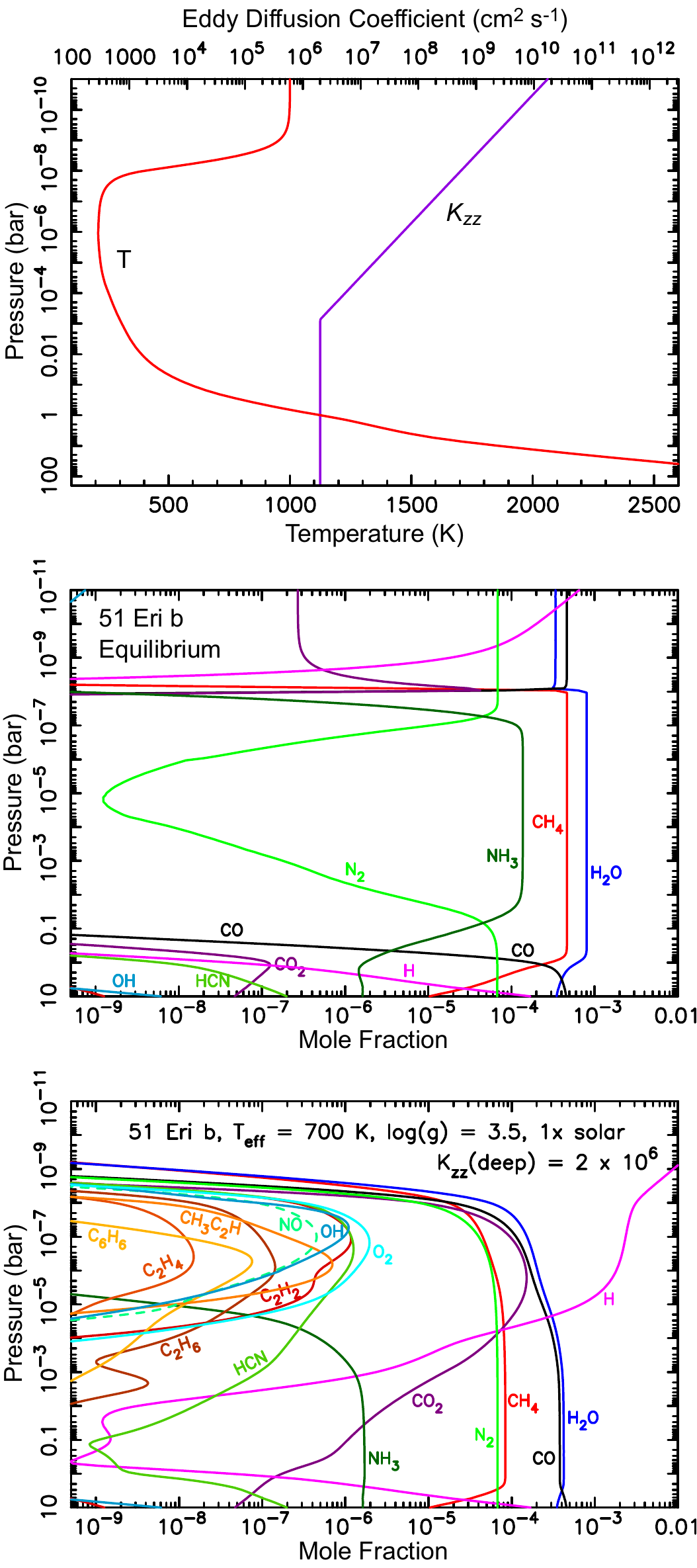}
\caption{Chemical model for 51 Eri b assuming \teff\ = 700 K, log($g$) = 3.5 cgs, mass = 2$M_J$, and solar metallicity: 
(Top) The temperature profile (red curve, bottom axis) from the radiative-convective equilibrium model of \citet{marley12} assuming 
the above bulk constraints, and the eddy diffusion coefficient profile (purple curve, top axis) adopted in the photochemical model; 
(Middle) the predicted thermochemical equilibrium mixing-ratio profiles for the major oxygen, carbon, and nitrogen species, as labeled, 
for the assumed pressure-temperature profile; (Bottom) mixing-ratio profiles predicted from our thermo/photochemical kinetics and 
transport model for the above thermal structure, \kzz\ profile, and assumed bulk elemental composition.  
A color version of this figure is available in the online journal.\label{51eri_700k}}
\end{figure}

\subsection{51 Eri b\label{sect51eri}}

51 Eridani b, a $\sim$20-Myr-old exoplanet that is cooler and closer to its star than HR 8799 b, was recently 
discovered with the Gemini Planet Imager \citep[GPI;][]{macintosh15}.  As with several other cool young Jupiters, the near-infrared 
flux and emission spectrum of 51 Eri b is difficult to reproduce theoretically without invoking cloudy or partial-cloud-covered 
atmospheres \citep{macintosh15}.  The spectra show evidence for strong methane and water absorption \citep{macintosh15}; however,
CH$_4$ is underabundant in comparison with chemical equilibrium, indicating that quenching is occurring and thus CO should also 
be abundant.  Model-data comparisons favor \teff\ = 700$\, ^{+50}_{-100}$ K, but the surface gravity is not well constrained 
\citep{macintosh15}.  Because the planet is colder, contains more quenched CH$_4$, and receives a stronger UV flux at its 
$\sim$14-AU orbital distance \citep{derosa15} than HR 8799 b, photochemistry is expected to be more important on 
51 Eri b, and indeed the recent independent photochemical modeling of \citet{zahnle16} demonstrates that this is the case.

Figure \ref{51eri_700k} shows the results for a 51 Eri b model with \teff\ = 700 K, log($g$) = 3.5 cgs (with assumed 
mass = 2$M_J$, radius $\approx$ 1.25$R_J$), \kdeep\ = 2$\scinot6.$ cm$^2$ 
s$^{-1}$, and a solar metallicity, with a thermal structure derived from the radiative-convective equilibrium model described 
in \citet{marley12}.  We added an arbitrary 1000-K thermosphere to the top of this model, in an analogy with Jupiter, but we found 
that the presence or absence of such a thermosphere has little effect on the results.  Note that this particular \kdeep\ value was selected 
because it produces a quenched CH$_4$ abundance consistent with the absorption depths seen the \citet{macintosh15} spectra.  Because 
the stratospheric temperature drops below 250 K, water recycling is relatively inefficient (see discussion in section \ref{sectgeneric} 
and in \citealt{zahnle16}), and as the H$_2$O becomes depleted due to photolysis, the production of CO$_2$ through CO + OH 
$\rightarrow$ CO$_2$ + H proceeds prolifically.  Carbon dioxide then becomes a major constituent on 51 Eri b at column abundances 
much greater than on HR 8799 b.  The inefficiency of water recycling also leads to greater abundances of other oxidized products 
such as O$_2$, NO, H$_2$CO, CH$_3$OH, and HNCO.  The high UV flux, large quenched CH$_4$ abundance, and cold stratosphere also allow 
greater production of complex hydrocarbons than in the HR 8799 b models, but again, none of the species in our models become 
abundant enough to condense to form hazes.  The predicted NH$_3$ abundance is significantly smaller than expected from chemical equilibrium due 
to the N$_2$-NH$_3$ quenching, and since N$_2$ is more stable chemically, the photochemical production of nitrogen species is limited 
by this relatively low NH$_3$ abundance.  HCN is the dominant product of the coupled carbon-nitrogen photochemistry, but with the 
low derived \kdeep\ for this model, quenching is less important in controlling the final HCN abundance than photochemistry.  The 
column abundances of several species from this model are provided in Table~\ref{tabcol51Eri}.

Although the disequilibrium composition of warmer young Jupiters like HR 8799 b resembles that of close-in hot Jupiters,
cooler young Jupiters like 51 Eri b are in a unique regime of their own.  Both photochemistry and quenching sculpt the 
composition, and the cooler stratospheric temperatures allow a variety of photochemical products to thrive.  Carbon 
dioxide becomes one of the dominant atmospheric constituents, in a process that is unique to cooler young Jupiters and 
brown dwarfs.  For stratospheres warmer than $\sim$250 K, the OH released from H$_2$O photolysis can still efficiently 
react with H$_2$ to recycle the water, but this reaction slows to a trickle at low temperatures.  A large percentage of 
the upper-stratospheric oxygen then is removed from CO and H$_2$O and ends up in CO$_2$.  This process does not occur on hot 
Jupiters because the temperatures are too high and the water and CO are efficiently recycled, and it does not occur on 
solar-system giant planets because overall stratospheric oxygen abundances are too low as a result of water condensation 
in the troposphere and small external oxygen influx rates due to interplanetary dust, cometary impacts, and satellite and 
ring debris \citep[e.g.,][]{moses04}.

\begin{deluxetable*}{llll}
\tabletypesize{\scriptsize}
\tablecaption{Column abundances for 51 Eri~\lowercase{b} models\label{tabcol51Eri}}
\tablewidth{400pt}
\tablecolumns{4}
\tablehead{
\colhead{ } & \colhead{Column abundance} & \colhead{Column abundance} & \colhead{Column abundance} \\
\colhead{Species} & \colhead{above 10 mbar} & \colhead{above 100 mbar} & \colhead{above 1 bar} \\
\colhead{ } & \colhead{(cm$^{-2}$)} & \colhead{(cm$^{-2}$)} & \colhead{(cm$^{-2}$)} 
}
\startdata
CH$_4$       & 6.8$\scinot19.$   & 6.8$\scinot20.$   & 6.8$\scinot21.$    \\
C$_2$H$_2$   & 4.8$\scinot14.$   & 4.8$\scinot14.$   & 4.8$\scinot14.$    \\
C$_2$H$_6$   & 1.6$\scinot15.$   & 3.6$\scinot15.$   & 1.2$\scinot16.$    \\
C$_3$H$_4$   & 1.5$\scinot14.$   & 1.5$\scinot14.$   & 1.5$\scinot14.$    \\
C$_6$H$_6$   & 3.7$\scinot14.$   & 7.9$\scinot14.$   & 1.4$\scinot15.$    \\
O$_2$        & 8.2$\scinot14.$   & 8.2$\scinot14.$   & 8.2$\scinot14.$    \\
H$_2$O       & 3.4$\scinot20.$   & 3.4$\scinot21.$   & 3.4$\scinot22.$    \\
CO           & 3.0$\scinot20.$   & 3.1$\scinot21.$   & 3.1$\scinot22.$    \\
CO$_2$       & 6.9$\scinot18.$   & 1.8$\scinot19.$   & 5.0$\scinot19.$    \\
H$_2$CO      & 2.4$\scinot13.$   & 1.8$\scinot14.$   & 7.4$\scinot15.$    \\
CH$_3$OH     & 6.8$\scinot13.$   & 8.7$\scinot14.$   & 2.4$\scinot15.$    \\
NH$_2$       & 7.5$\scinot14.$   & 7.5$\scinot14.$   & 7.6$\scinot14.$    \\
NH$_3$       & 1.2$\scinot18.$   & 1.3$\scinot19.$   & 1.4$\scinot20.$    \\
HCN          & 3.6$\scinot16.$   & 6.5$\scinot16.$   & 2.0$\scinot17.$    \\
HC$_3$N      & 3.6$\scinot14.$   & 4.9$\scinot14.$   & 4.9$\scinot14.$    \\
NO           & 5.1$\scinot13.$   & 5.1$\scinot13.$   & 5.1$\scinot13.$    \\
\enddata
\tablecomments{Model assumes log($g$) = 3.5 cgs.}
\end{deluxetable*}

\begin{figure}[htb]
\includegraphics[scale=0.66]{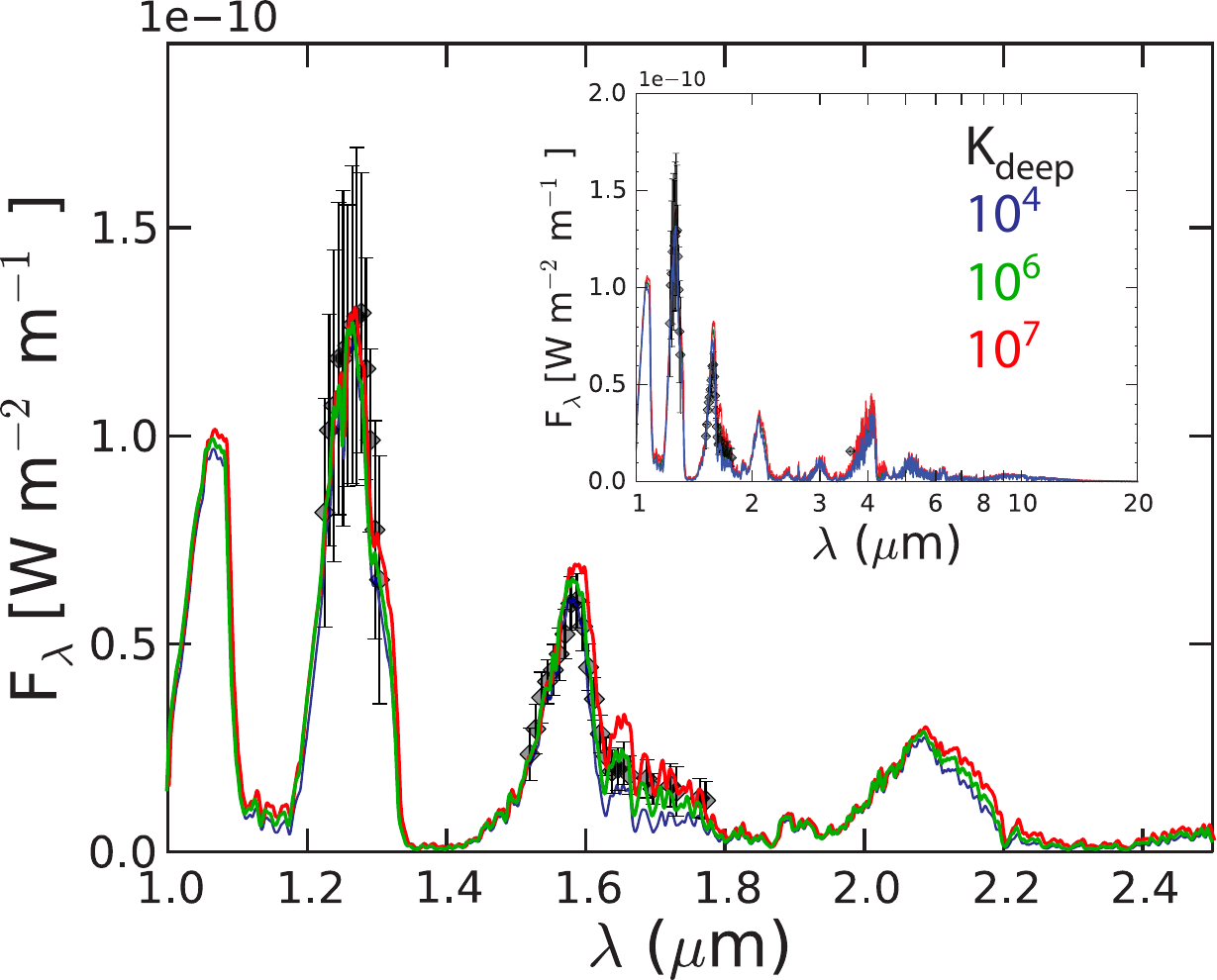}
\caption{The 51 Eri b GPI observations of \citet{macintosh15} (gray/black data points with error bars), 
in comparison with synthetic 
spectra from photochemical models that assume \teff\ = 700 K, $g$ = 3500 cm s$^{-2}$, a solar metallicity, $R$ = 0.8$R_J$, and 
\kdeep\ = 10$^4$ (blue), 10$^4$ (green), or 10$^7$ cm$^2$ s$^{-1}$ (red).  Lower \kdeep\ values lead to larger quenched CH$_4$ 
abundances and greater absorption in the long-wavelength side of the H band.  The inset shows an expanded wavelength range.  
As indicated by \citet{macintosh15}, we also find that we need to invoke partial cloud cover in order to reproduce 
the near-infrared observations.  For this particular analysis, we 
combined the spectrum of a cloud-free planet with one covered by a uniform global cloud, such that the ``cloud fraction'' was 
30\%.  The cloudy model assumed a uniform gray absorbing cloud with a base at $\sim$10 bar (representing Mg-silicates) and an 
optical depth of 1.76 between 0.1--1000 mbar (e.g., from Na$_2$S clouds or photochemical haze).  The planet was assumed to be 
29.4 pc from Earth.  A color version of this figure is available in the online journal.\label{51erispec}}
\end{figure}

Based on the column abundances predicted in this model (Table~\ref{tabcol51Eri}), CO$_2$ should be readily observable on 51 Eri b 
in the 4.2--4.3 $\mu$m and $\sim$15 $\mu$m regions, and perhaps even near 2.7--2.8 $\mu$m.  Carbon monoxide should also be 
observable at 2.3--2.4 $\mu$m (given sufficient spectral resolution) and at 4.5--4.9 $\mu$m.  Ammonia absorption is potentially 
detectable in bands near $\sim$1.5, 2, 3, and 6.15 $\mu$m if the lines can be disentangled from other absorbers and if the photosphere 
extends deep enough (i.e., is not obscured by high clouds), and NH$_3$ should be more readily detectable in the 9--11 $\mu$m region.
Figure \ref{51erispec} demonstrates that photochemical models with relatively large quenched methane abundances can reproduce the 
GPI spectra.

Our photochemical model results for 51 Eri b are qualitatively similar to those of \citet{zahnle16}, who use 
a different model, with a different numerical integrator, a different list of chemical species, different reaction 
rates and UV cross sections, different eddy diffusion coefficient profiles, and different assumptions about the stellar 
flux.  Comparisons between the two models therefore give some sense of the robustness of the theoretical predictions 
regarding the photochemical products.  The key quench reactions have different rate coefficients in the two models, 
so each model predicts slightly different quenched mixing ratios for any given \kdeep, but these differences are relatively
minor (i.e., within a factor of $\sim$2 for CH$_4$).  Both models predict strong photochemical production of CO$_2$ when 
the stratospheric temperatures fall below $\sim$250 K, and both models predict that a variety of complex hydocarbons 
will be produced photochemically in the middle and upper stratosphere of 51 Eri b.  The \citet{zahnle16} model includes 
sulfur chemistry, while the model presented here does not, and the list of complex hydrocarbon and nitrogen species 
and reactions is more extensive in the model here than in \citet{zahnle16}.  Because the nominal \citet{zahnle16} model 
does not contain as effective loss processes for some hydrocarbons such as C$_4$H$_2$, their model predicts greater 
abundances of these hydrocarbons, and \citet{zahnle16} presume that these hydrocarbons go on to inevitably
form PAHs and organic hazes, as an upper limit to possible photochemical haze formation.  The hydrocarbon chemistry 
presented here, which includes more C$_4$H$_x$, C$_5$H$_x$, and C$_6$H$_x$ 
hydrocarbons (including benzene) but not PAHs or heavier hydrocarbons, predicts more efficient conversion of the complex 
hydrocarbons back into simple hydrocarbons, CO, CO$_2$, and HCN, so the effectiveness of organic haze formation on 51 
Eri b has not been demonstrated.  Neither model includes ion chemistry or neutral-chemistry pathways to larger organic 
molecules.  The model here predicts a larger CO$_2$ abundance than is obtained in \citet{zahnle16}, for otherwise 
identical conditions.  This result seems to stem from the presence of sulfur species in the Zahnle et al.~model, 
where H$_2$S is readily destroyed by long-wavelength UV radiation, with the sulfur ending up in S$_8$ and oxidized 
species, and the excess H released in the process helping to convert some of the CO$_2$ back to CO.

\section{Discussion}

\subsection{Implications of disequilibrium $\COtwo$}

Our photochemical models for generic directly imaged planets and the specific young Jupiters HR 8799 b and 51 Eri b indicate 
that CO$_2$ is a major disequilibrium product on young Jupiters that is affected by both quenching and photochemistry.  The CO$_2$ 
abundance can increase significantly when stratospheric temperatures drop below $\sim$250 K, when metallicities are larger than 
solar, and when the eddy diffusion coefficients in the troposphere and lower stratosphere are relatively small (e.g., \kzz\ $<$ 10$^7$ cm$^2$
s$^{-1}$).  The CO$_2$ produced by disequilibrium processes is likely to affect the planet's emission spectrum, especially in the 
$\sim$4.3-  and 15-$\mu$m regions.  Detection could help constrain the planet's atmospheric metallicity, especially if \kzz\ at the quench 
point has already been constrained from the observed relative abundance of CH$_4$ and CO.  

Quenching (and potentially photochemistry, 
depending on local UV sources) will affect the CO$_2$ abundance on brown dwarfs, as well.  Brown dwarfs with lower \teff\ and 
colder stratospheres are expected to have more CO$_2$ simply as a result of quenching, and the CO$_2$ abundance can further be 
enhanced by photochemistry if there is a UV background sufficient to cause H$_2$O photolysis.  Perhaps galactic cosmic rays 
could also contribute to CO$_2$ production if that resulting chemistry leads to a similar destruction pathway for H$_2$O, and if 
OH + H$_2$ $\rightarrow$ H$_2$O + H is relatively inefficient (i.e., for cooler stratospheres).  If photochemistry or cosmic-ray 
chemistry can lead to CO$_2$ production on brown dwarfs, then that could explain the trends seen in the AKARI data of 
\citet{yamamura10}, who find that the CO$_2$ absorption band at $\sim$4.2--4.3 $\mu$m is enhanced tremendously in cooler 
late L and T dwarfs.  

\subsection{Implications of disequilibrium $\HCN$}

Hydrogen cyanide is the second most important product of disequilibrium chemistry on young Jupiters.  The HCN abundance is 
increased when the tropospheric \kdeep\ is large and the lower stratospheric \kzz\ is small (i.e., a stagnant lower stratosphere 
overlying a convective troposphere).  The strong HCN band near 3 $\mu$m may be detectable on young Jupiters if high clouds do not 
fully obscure the upper troposphere, although a relatively high spectral resolution may be needed to disentangle the HCN lines from 
other absorbers such as CH$_4$.  A source of atomic H from H$_2$S and PH$_3$ at depth (not included in this model) could lead to 
increased HCN abundances by attacking CH$_4$ and NH$_3$ to produce CH$_3$ and NH$_2$, augmenting coupled carbon-nitrogen photochemistry 
through CH$_3$NH$_2$ pathways such as scheme (6) above and others described more fully in \citet{moses10,moses11}.

\subsection{Implications for hazes}

Our neutral carbon, oxygen, and nitrogen photochemistry described here does not lead to the production of organic hazes in 
our young-Jupiter models.  Some complex organics are produced in the models, but the abundances are not large enough in these generally 
warm stratospheres to lead to supersaturations.  Note that the complex organics in our directly imaged planet models are less 
abundant than on Jupiter and Saturn, and yet the stratospheric hazes on our solar-system giant planets are not optically 
thick when the refractory organics such as C$_4$H$_2$, C$_4$H$_{10}$ and C$_6$H$_6$ become supersaturated and condense
\citep[e.g.][]{moses04,west04}.  Therefore, optically thick hydrocarbon hazes on young Jupiters might not be expected.  However, 
ion chemistry in the auroral regions of Jupiter and Saturn seems to be more effective at producing polycyclic aromatic 
hydrocarbons (PAHs) and other complex hydrocarbons that then condense in the high-latitude stratosphere to form thicker 
``polar hoods'' of aerosols \citep[e.g.,][]{pryor91,wong00,wong03,friedson02}.  

Ion chemistry on young Jupiters may therefore enhance the production of complex hydrocarbons and eventual hazes, but 
even in the presence of ionization, optically thick haze formation is not guaranteed.  For example, solar ionization 
of hydrocarbons is effective at low-to-mid latitudes on Jupiter and Saturn \citep[e.g.,][]{kim94,kim14}, but optically 
thick stratospheric photochemical hazes do not result from this process.  Several Titan laboratory simulations demonstrate 
that PAH formation is favored when molecular nitrogen is present and is ionized \citep[e.g.,][]{imanaka07}.  Whether this 
rich Titan-like ion chemistry can still occur in warmer H$_2$-dominated situations, where the CH$_4$ homopause limits 
altitude range over which the ion chemistry is effective and for which O and OH are present to potentially short-circuit 
the process by oxidizing the carbon and sending it preferentially to CO and CO$_2$, remains to be seen.  Laboratory 
investigations similar to those of \citet{imanaka09}, \citet{sciamma10}, \citet{peng13}, and \citet{horst14} but that 
are specifically designed for stratospheric conditions on young Jupiters would further our understanding of the likelihood 
of organic photochemical hazes.  

Other possibilities for clouds and hazes on young Jupiters include the standard equilibrium cloud sequence 
\citep[e.g.,][]{morley12,marley13}, for which magnesium-silicate clouds might affect spectra if they are vertically 
thick,  and for which Na$_2$S clouds are likely to reside within the photospheres of many young Jupiters (see Fig.~\ref{figtemp}).  
\citet{zahnle16} identify elemental sulfur as another intriguing possible photochemical haze that is particularly 
likely when the stratosphere is relatively cold and well irradiated.  Hydrogen sulfide is chemically fragile, and 
although the kinetics of sulfur species is not well determined for relevant atmospheric conditions, the 
formation of S$_8$ molecules as described by \citet{zahnle16} seems a likely possibility.  \citet{zahnle16} find that 
sulfur chemistry would destroy all photospheric H$_2$S for a 51 Eri b planet at an orbital distance of $\lta$ 600 AU.  
Phosphine (PH$_3$) is also a chemically fragile molecule, and the phosphorus may end up in elemental phosphorus or 
other relatively refractory photochemical species that could eventually form hazes.  The identity of the clouds that 
seem to affect the spectra of young Jupiters is therefore unclear, but there are many candidate materials, including 
photochemical hazes.

\subsection{Implications for JWST and WFIRST}

The \textit{James Webb Space Telescope} (JWST) will have the capability of obtaining high-contrast ($\geq10^{-7}$) 
images of young giant planets \citep{beichman10,clampin11}, with well-separated ones (3$''$-10$''$) being 
amenable to spectroscopic characterization.  The Near Infrared Camera (NIRCam), Mid-Infrared Instrument (MIRI), 
and Near-InfraRed Imager and Slitless Spectrograph (NIRISS) all have imaging modes, with MIRI and NIRCam having 
coronagraph options \citep[e.g.,][]{krist07,boccaletti15,debes15} and NIRISS having an Aperture Masking 
Interferometry (AMI) mode \citep{artigau14}.  Additionally, the JWST Near Infrared Spectrograph (NIRSpec) and 
MIRI instruments both have Integral Field Units (IFUs) \citep{arribas07,wells15} that will allow for 
medium to high resolution ($R\sim1000-3000$) spectra of planets with angular separations greater than 3$''$.
Collectively these instruments can provide high-sensitivity photometry and spectroscopy of giant planets 
orbiting at a variety of separations (0.1$''$--10$''$) from their host stars, 
over the wavelength range $\sim$0.6--28 $\mu$m \citep[see the review of][]{beichman10}.  JWST will therefore 
provide the opportunity for detecting the excess CO$_2$ absorption we predict here near $\sim$4.3 $\mu$m 
(see Fig.~\ref{specgenericco2}) and $\sim$15 $\mu$m, as well as provide better 
constraints on the quenched methane abundance from absorption in the 3.3- and 7.7 $\mu$m bands, quenched NH$_3$ from the 
10--11 $\mu$m band, signatures of photochemically produced species such as HCN (near $\sim$3, 7, and 14 $\mu$m), C$_2$H$_2$ 
(at 13.6 $\mu$m), and C$_2$H$_6$ (near 12.2 $\mu$m).  

In the next decade, NASA's Wide Field Infrared Survey Telescope (WFIRST), equipped with an optical (0.4-0.95~$\mu$m)
Coronagraphic Instrument (CGI), will obtain photometry and spectra for extrasolar planets with
contrasts as low as 10$^{-10}$ and angular separations between 0.1$''$ and 0.5$''$ \citep{traub16}.
The WFIRST CGI prime survey will target a number of planets detected via radial velocity both photometrically
(0.4-0.95~$\mu$m) and spectroscopically (0.6-0.95~$\mu$m).  These planets will be relatively cool, mature 
giant planets orbiting closer to their stars than Jupiter.  Many of these planets will be warm 
enough to lack ammonia, hydrogen sulfide, and in some cases, water clouds \citep{marley14}.  Thus their predicted 
photospheric equilibrium chemistry is comparable to the directly imaged planets studied here, although their deep 
atmospheres will be colder.  Methane is likely the dominant form of carbon in the atmospheres of these planets, and 
although detailed predictions for the photochemistry of such worlds awaits future studies, the implications of our models 
presented here point to a likely rich photochemistry.  Preliminary investigations \citep{sharp04} suggest that the older 
Jupiters with neither water nor ammonia trapped in tropospheric clouds \citep[i.e., the ``Class III'' planets in the 
terminology of][]{sudarsky03} will have stratospheric chemistry similar to what is described for our young-Jupiter models.  
\citet{sharp04} find that hydrocarbon and nitrile photochemistry is even more prevalent on ``Class II'' planets, for which 
water condenses in the troposphere but ammonia does not \citep{sudarsky03}, leading to high production rates for HCN, 
CH$_3$CN, other nitriles, and complex hydrocarbons.  Coupled C$_2$H$_2$-NH$_3$ photochemistry will likely produce 
high-molecular-weight organo-nitrogen compounds \citep{moses10} and polycyclic aromatic hydrocarbons.  Photochemical 
production of hazes will likely be important, which can sculpt the ultraviolet and blue reflection spectra 
\citep[e.g.,][]{griffith98}, as on our own solar-system giant planets, thereby affecting the reflection photometry 
or signatures observed by WFIRST.

\section{Conclusions}

Our modeling of directly imaged exoplanets indicates that the atmospheric composition of these young Jupiters is 
expected to be far from chemical equilibrium, confirming the results of previous studies that indicate CH$_4$ and CO 
quenching is occurring on these planets
\citep[e.g.,][]{bowler10,hinz10,janson10,janson13,barman11hr8799,barman112mass,barman15,galicher11,marley12,skemer12,skemer14,ingraham14,currie14,zahnle14}.
Transport-induced quenching will cause CO, and not CH$_4$, to be the dominant carbon constituent on most lower-gravity 
young Jupiters with \teff\ $\ge$ 600 K, for all reasonable estimates of the strength of deep-atmospheric convection.  This 
conclusion is inevitable.  The first line of attack for interpreting young-Jupiter spectra should therefore always be 
models that include quenching.  Photochemistry can also play a significant role in young-Jupiter atmospheres, especially 
on cooler planets that receive strong ultraviolet irradiation from their host stars.  

Rapid transport in the deep atmosphere also leads to quenching of H$_2$O at the same point as the quenching of 
CO and CH$_4$.  This effect does not appear to be as widely realized as the CH$_4$--CO quenching phenomenon, but it is 
important, as the quenching can occur in a region where the equilibrium H$_2$O mixing ratio is increasing with altitude, 
with quenching then causing a lower-than-expected H$_2$O abundance on young Jupiters.  In this situation, the 
oxygen is preferentially tied up in quenched CO rather than H$_2$O, and the water mixing ratio can be a factor of a few 
lower than equilibrium predictions.  Water is the dominant infrared opacity source on young Jupiters, and the 
fact that quenching can alter the expected abundance can in turn affect the predicted thermal structure, cooling history, 
spectral energy distribution, and inferred C/O ratio of these planets (the latter due to the fact that the CO abundance 
is typically difficult to constrain precisely).  Models that consider the thermal evolution of giant planets or that 
predict the current thermal structure of young Jupiters should take the quenching of H$_2$O into account, although 
this factor is not likely to have as large an impact as clouds or initial conditions.

Quenching will also affect the relative abundances of NH$_3$ and N$_2$, favoring N$_2$ rather than NH$_3$ at 
the quench point.  Although NH$_3$ is not expected to be the dominant nitrogen-bearing constituent, the quenched 
ammonia abundance may still be observable on young Jupiters if the photosphere extends into the upper troposphere 
and is not obscured by clouds.  The quenched NH$_3$ mixing ratio increases as \teff\ decreases.

Other potentially observable constituents that are expected to be negligible in equilibrium models but that are 
affected by disequilibrium chemical processes include CO$_2$ and HCN.  These molecules are affected by both 
quenching and photochemistry.  The quenching process leads to increases in the HCN abundance when deep atmospheric 
mixing is strong, while increases in CO$_2$ are favored when deep atmospheric mixing is weak.  Photochemical 
production of both HCN and CO$_2$ is more important for weak lower-stratospheric mixing and strong UV irradiation.
Effective temperatures of 900--1400 K favor larger HCN column abundances, whereas the CO$_2$ column abundance 
increases significantly for lower \teff, and specifically for low stratospheric temperatures $T$ $\lta$ 250 K.  
When stratospheric temperatures are low, the reaction OH + $\Htwo$ $\rightarrow$ $\HtwoO$ + H becomes ineffective, 
and OH + CO $\rightarrow$ $\COtwo$ + H can compete \citep[see also][]{zahnle16}, depleting the upper stratospheric 
H$_2$O and CO, and significantly increasing the column abundance of photochemically produced CO$_2$.  On cooler 
planets like 51 Eri b, the CO$_2$ peak mixing ratio can even exceed that of CH$_4$ and rivals that of CO and H$_2$O 
in the upper stratosphere.  Carbon dioxide is likely to be observable on all young Jupiters with moderate-to-low 
atmospheric mixing, but will be especially important on cooler planets.  Hydrogen cyanide is less likely to be 
observable on young Jupiters, but it may be detectable in the $\sim$3 $\mu$m band given favorable atmospheric 
conditions (including the absence of high clouds) and sufficient spectral resolution to disentangle the lines from 
other nearby absorbers.

Complex hydrocarbons also form photochemically on young Jupiters, but the generally warm stratospheric temperatures 
and large H abundance make them less stable than on the giant planets in our solar system.  Oxidation of the carbon 
to form CO and CO$_2$ also competes effectively, unlike on our own giant planets.  It is unlikely that hydrocarbons 
produced from neutral photochemistry will be observable on young Jupiters.  Note that the models presented here include 
only H-, C-, O-, and N-bearing species.  Ion chemistry is not included, nor is the neutral photochemistry of other 
volatiles like sulfur and phosphorus.  As shown by \citet{zahnle16}, sulfur chemistry can alter some of the predictions 
regarding the abundances of C-, N-, and O- species.  Although organic hazes do not form from 
the neutral chemistry considered here, ion chemistry may augment the production of refractory organics, as on Titan and 
in the auroral regions of Jupiter \citep[e.g.,][]{wong00,waite07,vuitton07}.  Future laboratory and theoretical modeling 
should focus on this possibility.  Laboratory studies that investigate the kinetics of C$_3$H$_2$ and C$_3$H$_3$ reactions 
with other hydrocarbon radicals and molecules would aid exoplanet photochemistry studies.  Other possible photochemically 
produced hazes include elemental sulfur \citep{zahnle16}, elemental phosphorus or other refractory phosphorus species, 
and refractory products from coupled C$_2$H$_2$--NH$_3$ chemistry \citep[e.g.,][]{ferris88,keane96,moses10}.

Detection and abundance determinations for key molecules like CH$_4$, H$_2$O, CO, CO$_2$, and NH$_3$ can help constrain 
planetary properties and potentially break other modeling degeneracies.  The CH$_4$ and NH$_3$ mixing ratios are strong 
indicators of the strength of deep atmospheric mixing, \kdeep, as well as the planet's effective temperature, \teff.  Simultaneous 
measurements of the abundance of H$_2$O and CO can provide additional constraints on \teff, surface gravity, and metallicity.  
The CO$_2$ abundance is very sensitive to metallicity \citep[e.g.,][]{lodders02,moses13gj436}, and can also become quite large 
for low \teff, low stratospheric \kzz, and high UV irradiance. 

%
The disequilibrium composition of warmer young Jupiters (i.e., \teff\ $\gta$ 900 K), such as HR 8799 b, 
resembles that of close-in transiting giant planets.  Transport-induced quenching is the dominant process 
driving the atmospheres out of equilibrium, and the stratospheres are too warm to allow many of the photochemical 
products to survive, other than molecules with strong bonds like C$_2$H$_2$ and HCN.  However, cooler young Jupiters 
(\teff\ $\lta$ 700 K) like 51 Eri b can have a rich and interesting photochemistry that differs distinctly from that 
of either hot Jupiters or the cold giant planets in our solar system.  The quenched abundances of photochemically 
active CH$_4$ and NH$_3$ tend to be greater for lower \teff, and hydrocarbon photochemical products survive more readily 
when stratospheric temperatures are low.  Oxidation of the carbon and nitrogen species can also proceed much more 
effectively when stratospheric temperatures are low (due to a reduction in efficiency of H$_2$O recycling), leading to 
oxidized products like NO, O$_2$, and especially CO$_2$.  Carbon dioxide is likely to be a major absorber on cooler 
young Jupiters.

Cooler directly imaged giant planets that receive moderate-to-high UV flux from their host stars fall into a unique and 
interesting chemical regime that is controlled by both transport-induced quenching and an active, rich photochemistry.  
This chemical regime has no representation in our own solar system because the terrestrial planets have very different 
atmospheric compositions and the colder giant planets have key oxygen and nitrogen species tied up in condensates at 
depth, so that coupled nitrogen-carbon, oxygen-carbon, and nitrogen-oxygen photochemistry is suppressed.  The simultaneous 
presence of H$_2$O, CO, CH$_4$, N$_2$, and NH$_3$ on cooler young Jupiters leads to complex photochemical interactions 
with both oxidized and reduced products being stable, and small amounts of high-molecular-weight pre-biotic molecules 
being able to form and survive.  With dedicated ground-based campaigns ramping up their search for young
Jupiters \citep[e.g.,][]{macintosh15,vigan16,tamura16}, and missions such as JWST, GAIA, and WFIRST gearing up or 
being planned, we look forward to many future reports of the atmospheric composition of directly imaged giant planets.



\acknowledgments

This material is based upon work supported by the National Aeronautics and Space Administration through 
NASA Exoplanet Research Program grant NNX15AN82G (initially) and NNX16AC64G.
We thank Kevin France for useful advice on constructing the stellar ultraviolet fluxes, and the anonymous reviewer 
for a thorough review of the manuscript.  Portions of the stellar spectra were compiled using data from the Mikulski 
Archive for Space Telescopes (MAST) at STSci and the X-exoplanet archive at the CAB.





\bibliographystyle{aasjournal}
\bibliography{references}


\clearpage

\end{document}